\documentclass[%
 reprint,
 amsmath,amssymb,
prx,
floatfix,
superscriptaddress
]{revtex4-2}

\usepackage{mathtools}
\usepackage{graphicx}
\usepackage{dcolumn}
\usepackage{bm}

\usepackage[caption=false]{subfig}

\usepackage{algorithm}
\usepackage[noend]{algpseudocode}
\algrenewcommand\algorithmicdo{}

\makeatletter
\renewcommand{\ALG@name}{Procedure}
\makeatother

\usepackage[linktocpage=true,
  colorlinks=true, 
  pdfborder={0 0 0},
  linkcolor=blue,
  citecolor=red,
  filecolor=yellow,
  urlcolor=blue,
  bookmarks,
  pdfauthor={},
]{hyperref}

\usepackage{orcidlink}

\usepackage{ifthen}
\newcounter{is_qcircuit_used}
\newcounter{are_figs_merged}
\newcommand{\argmin}{\mathop{\rm arg~min}\limits}

\begin{document}

\preprint{APS/123-QED}

\title{
Error-corrected phase estimation averaged over variable grids on\\a trapped-ion quantum computer:\\hyperacuity spectra of a CO molecule adsorbed onto $\chi$-Fe$_5$C$_2$ 
}

\author{Taichi Kosugi\orcidlink{0000-0003-3379-3361}}
\email{kosugi.taichi@gmail.com}
\affiliation{
Quemix Inc.,
Taiyo Life Nihombashi Building,
2-11-2,
Nihombashi Chuo-ku, 
Tokyo 103-0027,
Japan
}

\affiliation{
Department of Physics,
The University of Tokyo,
Tokyo 113-0033,
Japan
}

\author{Hirofumi Nishi\orcidlink{0000-0001-5155-6605}}
\affiliation{
Quemix Inc.,
Taiyo Life Nihombashi Building,
2-11-2,
Nihombashi Chuo-ku, 
Tokyo 103-0027,
Japan
}

\affiliation{
Department of Physics,
The University of Tokyo,
Tokyo 113-0033,
Japan
}

\author{Keito Kasebayashi}
\affiliation{
Business Creation Sector R\&D Center,
MITSUI KINZOKU COMPANY,
LIMITED,
1333-2 Haraichi,
Ageo-shi,
Saitama,
362-0021,
Japan
}

\author{Hiroki Takahashi\orcidlink{0009-0006-5086-700X}}
\affiliation{
Business Creation Sector R\&D Center,
MITSUI KINZOKU COMPANY,
LIMITED,
1333-2 Haraichi,
Ageo-shi,
Saitama,
362-0021,
Japan
}

\author{Yu-ichiro Matsushita\orcidlink{0000-0002-9254-5918}}
\affiliation{
Quemix Inc.,
Taiyo Life Nihombashi Building,
2-11-2,
Nihombashi Chuo-ku, 
Tokyo 103-0027,
Japan
}

\affiliation{Quantum Materials and Applications Research Center,
National Institutes for Quantum Science and Technology (QST),
2-12-1 Ookayama, Meguro-ku, Tokyo 152-8550, Japan
}

\affiliation{
Department of Physics,
The University of Tokyo,
Tokyo 113-0033,
Japan
}

\date{\today}

\begin{abstract}
Quantum phase estimation (QPE) is an underlying technology for extracting the excitation spectra of many-electron systems,
yet its practical use on current hardware is hindered by low grid resolution and environmental noises.
Here we propose QPE averaged over variable grids (QAVG),
a vernier-type approach that combines low-resolution QPE with multiple origin shifts and physically motivated continuous parametrization to reconstruct the spectra accurately.
We introduce this approach into an end-to-end workflow for the {\it ab initio}-based model system for a CO molecule adsorbed onto the $\chi$-Fe$_5$C$_2$ surface.
We perform experiments on Quantinuum H2-2 using both physical QPE circuits and logical QPE circuits encoded in the Steane code with offline bit-flip correction.
We demonstrate that QAVG accurately reconstructs the spectra with deviations much smaller than the nominal QPE resolution, even when the noisy histograms are used.
The cost landscapes averaged over the shifted grids substantially suppress the local minima arising from the spectral leakage,
thereby stabilizing the optimization of trial parameters.
These results indicate that QAVG provides a robust route to quantum simulations of correlated spectra toward the era of early-fault-tolerant quantum computers.
\end{abstract}

\maketitle 

\section{Introduction}
\label{sec:introduction}

Quantum algorithms that are expected to outperform classical ones have been proposed in various scientific fields.
Many of these algorithms are designed as combinations of a small number of elementary quantum subroutines that serve as foundational building blocks.
One of the most prominent among them is quantum phase estimation (QPE) \cite{Nielsen_and_Chuang,bib:7441_ref_43,bib:4724,bib:7441_ref_68,bib:7441_ref_69,bib:4743,bib:6827,bib:6807,bib:7458},
variants of which basically extract the eigenvalues of an implementable unitary operator (or an implementable exponentiated Hermitian operator) from an input many-qubit state overlapping with the corresponding eigenstates.
QPE is particularly efficient for applications in quantum chemistry,
where it is known to enable the acquisition of physical quantities such as one-particle Green’s functions (GFs) \cite{bib:5005} and linear-response functions \cite{bib:5163, bib:6602} in second-quantized formalism more efficiently than classical algorithms.
It is because such QPE sampling does not require diagonalization for the huge Hilbert subspaces where the excited states reside as long as the initial state is already prepared.
These quantities provide essential information for understanding the static and dynamic properties particularly in systems where correlation effects play significant roles.
To perform such sophisticated algorithms with high precision,
it is essential to implement them in a fault-tolerant (FT) manner \cite{Nielsen_and_Chuang} based on logical states by using high-fidelity physical qubits and appropriate error-detecting or error-correcting codes.

Given the recent development of quantum hardware \cite{bib:6326, bib:7108, bib:7111, bib:7246, bib:7301},
much more attention is paid to practical applications of quantum computation based on logical qubits than ever.
We find such works that used real quantum computers as follows.
Yamamoto et al.~\cite{bib:6484} performed
the Bayesian QPE for estimating the ground state energy of a hydrogen molecule  based on the $[[6,4,2]]$ error-detecting code,
which is a member of the family of the $[[k+2,k,2]]$ Iceberg codes \cite{bib:6473}, using a trapped-ion quantum computer.
Zhong et al.~\cite{bib:7344} performed the variational quantum eigensolver (VQE) experiments on a superconducting quantum computer to prepare the ground state of a hydrogen molecule,
where they employed a hybrid protocol that integrates Pauli twirling and probabilistic error cancellation based on the $[[4,2,2]]$ code \cite{bib:5305,bib:7452}.
Nishi et al.~\cite{bib:6948} performed
the probabilistic imaginary-time evolution (PITE) \cite{bib:5409,bib:5737,bib:6103,bib:6231,bib:6242,bib:6571,bib:6630,bib:6805,bib:7453,Kobori_penalty} based on the $[[6,4,2]]$ code using a trapped-ion quantum computer to prepare the ground and excited states of NV center models for promising materials as spin qubits.
He et al.~\cite{bib:7343} carried out experiments of the quantum approximation optimization algorithm (QAOA) based on the Iceberg codes up to 20 logical qubits by using a trapped-ion quantum computer.
Nishi et al.~\cite{bib:7441} performed QPE sampling for the excited states of FePO$_4$ based on the $[[6,4,2]]$ code using a trapped-ion quantum computer to obtain the x-ray absorption spectroscopy (XAS) spectra.
Yamamoto et al.~\cite{bib:7290} performed QPE for a hydrogen molecule based on the $[[7,1,3]]$ error-correcting code, also known as the Steane code \cite{bib:5281}, using a trapped-ion quantum computer to estimate the ground state energy by maximizing the likelihood.
It is noted here that Mayer et al.~\cite{bib:6542} benchmarked the logical quantum Fourier transform (QFT),
which will be undoubtedly one of the most important elementary technologies in the FT quantum computation (FTQC) era,
for three logical qubits in the Steane code by using a trapped-ion quantum computer.
They adopted recursive gate teleportation \cite{bib:7292, bib:7293, bib:7294, bib:6542} for implementing logical non-Clifford operations.

\begin{figure*}
\begin{center}
\includegraphics[width=17cm]{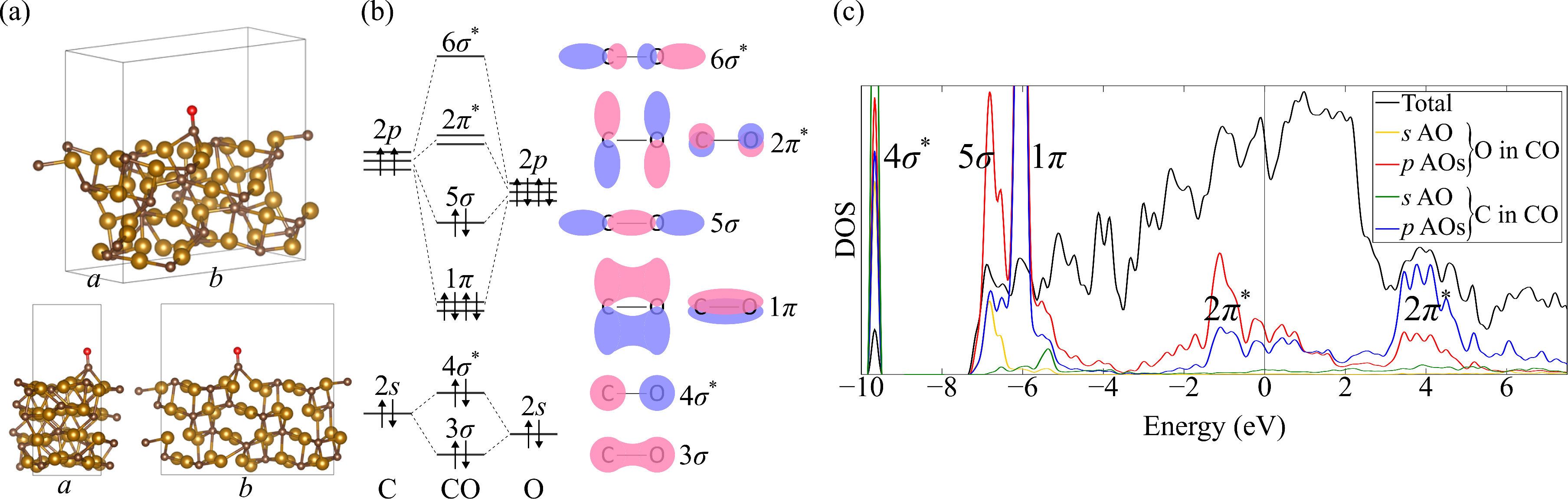}
\end{center}
\caption{
(a)
Unit cell used for DFT calculations in the present study.
This periodic system mimics the non-periodic adsorption system,
where a CO molecule is adsorbed onto the (510) surface of $\chi$-Fe$_5$C$_2$.
This figure was drawn by using VESTA \cite{VESTA}.
(b)
Schematic energy diagram of the molecular orbitals (MOs) of an isolated CO molecule.
The amplitudes of the MOs are also depicted.
(c)
Electronic density of states (DOS) obtained in the DFT calculation as functions of an energy measured from the Fermi level.
The total DOS is shown as black curves.
The projected DOS onto the atomic orbitals (AOs) at the CO molecule are also shown,
where the MOs giving rise to the major contributions are written. 
The total DOS in the figure has been multiplied by a factor since the raw values are very large compared to the projected DOS.
}
\label{fig:Fe5C2-CO_cell_and_dft_dos}
\end{figure*}

Iron-based catalysts in Fischer--Tropsch synthesis (FTS) are known to play crucial roles in chemical industry. 
Such catalysts actually undergo the carburization process to form various iron carbides.
Amongst them, $\chi$-Fe$_5$C$_2,$ also known as H\"agg carbide, is particularly interesting
since its surface with a high Miller index (510) has been experimentally detected \cite{bib:7429_ref_14, bib:7429_ref_26, bib:7429_ref_27} and
the calculations predicted its high thermodynamic stability compared to other facets \cite{bib:7429}.
In addition, CO activation on the surfaces of $\chi$-Fe$_5$C$_2$ is important for the initiation of the FTS. 
Realistic elucidation and description of the activation process for a CO molecule on transition metal surfaces \cite{bib:7430,bib:7431} from a viewpoint of electronic theory demands the incorporation of electronic correlation, as discussed later.
We therefore adopt an end-to-end workflow in the present study for $\chi$-Fe$_5$C$_2$ where a CO molecule is adsorbed onto the (510) surface, as depicted in Figs.~\ref{fig:Fe5C2-CO_cell_and_dft_dos}(a) and (b).
The workflow starts from electronic-structure calculations based on the density functional theory (DFT),
from which the maximally localized Wannier orbitals (MLWOs) \cite{bib:4596} essential for the adsorption are constructed on a classical computer.
The second-quantized model Hamiltonian for them is then transformed into qubit representation.

For obtaining the one-particle GF of the model system,
we propose QPE averaged over variable grids (QAVG),
which we apply to physical and logical qubits on a trapped-ion quantum computer.
This approach enables us to determine the physical quantities based on observed histograms for the QPE circuits characterized by shifts of energy origin smaller than the grid resolution.
The philosophy of QAVG reminds us of the hyperacuity (see, e.g., Refs.~\cite{bib:7461,bib:7462}),
that achieves rather fine spatial resolution compared to the distance between the centers of photoreceptors on the retina by making use of multiple neural information sources.
Although that is just an analogy,
the QAVG cost function in fact has much fewer local minima than that for a single origin shift and it thus facilitates the parameter optimization, as will be demonstrated later.

\section{Methods}

\subsection{DFT calculations and MLWOs}

\subsubsection{MLWOs for adsorption system}

For examining the electronic states of the adsorbed CO molecule,
we focus on the CO molecule as the adsorbate and the three Fe atoms as the adsorbent directly bonded to the molecule.
We construct the MLWOs from the Bloch wave functions obtained from DFT calculations for the subsequent many-electron approaches on classical and quantum computers.
The transfer integrals between the MLWOs are obtained from the standard Wannierization machinery,
after which we calculate the screened Coulombic repulsion energies based on the constrained random-phase approximation (cRPA) approach \cite{bib:7438}.

\subsubsection{Dimer model from MLWOs}

Given the capabilities of current quantum hardware realized so far,
we decide to examine a small model composed of two electronic orbitals,
instead of the original five MLWOs.
We construct the model from $p$- and $d$-type orbitals based on the MLWOs and define the second-quantized Hamiltonian
$
\mathcal{H}^{(\mathrm{dimer})}
=
\mathcal{H}_1^{(\mathrm{dimer})}
+
\mathcal{H}_2^{(\mathrm{dimer})}
,
$
where
\begin{gather}
    \mathcal{H}_1^{(\mathrm{dimer})}
    \nonumber \\
    =
        \sum_{\sigma = \uparrow, \downarrow}
        \left(
            \sum_{\kappa = p, d}
            (\varepsilon_\kappa - \Delta \mu )
            n_{\kappa \sigma}
            +
            t_{pd}
            (
                a_{d \sigma}^\dagger
                a_{p \sigma}
                +
                \mathrm{H.c.}
            )
        \right)
\end{gather}
is the one-body part and
\begin{align}
    \mathcal{H}_2^{(\mathrm{dimer})}
    =
            \sum_{\kappa = p, d}
            U_\kappa
            n_{\kappa \uparrow}
            n_{\kappa \downarrow}
\end{align}
is the two-body part.
$\varepsilon_\kappa$ is the orbital energy of type $\kappa.$
$t_{p d}$ is the transfer integral.
$U_\kappa$ is the on-site Coulomb repulsion energy between two electrons of type $\kappa.$
$a_{\kappa \sigma}^\dagger$ and $a_{\kappa \sigma}$ are the creation and annihilation operators of an electron, respectively, that satisfy the fermionic anticommutation relation.
$n_{\kappa \sigma} \equiv a_{\kappa \sigma}^\dagger a_{\kappa \sigma}$ is the number operator.
We take the $z$ direction as the spin quantization axis.
This model is mathematically nothing but a Hubbard dimer.
Although the DFT Fermi level has already been taken into account as the orbital energies in the dimer,
we introduce a shift $\Delta \mu$ of chemical potential artificially so that the ground state of the dimer over all the possible electron numbers is a two-electron state.
It is because the $2 \pi^*$-derived states in the DFT calculation are roughly half filled [see Fig.~\ref{fig:Fe5C2-CO_cell_and_dft_dos}(c)] and we hope the dimer model to reflect this filling.
In fact, the ground state of the dimer model constructed in this way is found to belong to the sector of two electrons $(n_e = 2)$ and no spin polarization $(S_z = 0),$ as discussed later.

\subsection{Single-qubit models}

\subsubsection{Dimension reduction for quantum computation}

The original protocol for obtaining the one-particle GF at a zero temperature via QPE sampling \cite{bib:5005} begins with the preparation of the ground state $| \Psi_{\mathrm{gs}} \rangle,$
which is fed into the probabilistic-excitation circuit and finally into the QPE circuit.
If we map the many-electron states of the dimer to many-qubit states naively,
four qubits are necessary.
The implementation for such mapping, however, leads inevitably to a rather deep circuit due to the controlled Pauli operators in the excitation part and
the controlled real-time evolution (RTE) operators in the QPE part.
Since the primary purpose of the present work is to demonstrate the applicability of the QAVG approach by performing it for logical qubits,
we choose to start from preparing deterministically the excited states instead of the probabilistic excitations.
This approach allows us to reduce the dimension of subspace for the involved electronic states to two, as described below.

\subsubsection{Electron-excited states}

The electron-excited states
$a_{p \uparrow}^\dagger | \Psi_{\mathrm{gs}} \rangle$ and
$a_{d \uparrow}^\dagger | \Psi_{\mathrm{gs}} \rangle$
belong to the two-dimensional subspace for $n_e = 3$ and $S_z = 1/2,$
which is spanned by
$
| 0; + \uparrow \rangle
\equiv
a_{p \uparrow}^\dagger
a_{d \uparrow}^\dagger
a_{p \downarrow}^\dagger
| \mathrm{vac} \rangle
$
and
$
| 1; + \uparrow \rangle
\equiv
a_{p \uparrow}^\dagger
a_{d \uparrow}^\dagger
a_{d \downarrow}^\dagger
| \mathrm{vac} \rangle
.
$
$| \mathrm{vac} \rangle$ is the vacuum state of the electron field.
The Hamiltonian matrix for this subspace is expressed as
\begin{align}
    \mathcal{H}^{(+ \uparrow)}
    =
        \begin{pmatrix}
            2 \varepsilon_p + \varepsilon_d - 3 \Delta \mu + U_p & t_{p d} \\
            t_{p d} & \varepsilon_p + 2 \varepsilon_d - 3 \Delta \mu + U_d
        \end{pmatrix}
        .
    \label{Hamiltonian_up_created}
\end{align}
We map the two electronic states 
$| 0; + \uparrow \rangle$ and $| 1; + \uparrow \rangle$ to the single-qubit states
$| 0 \rangle$ and $| 1 \rangle$, respectively, when performing QPE for the spin-up-created states on a quantum computer.
The electronic Hamiltonian for the excited states accordingly leads the qubit Hamiltonian of the form
$
\mathcal{H}^{(+ \uparrow)}_q
=
h_0 I + h_x \sigma_x + h_z \sigma_z.
$

\subsubsection{Hole-excited states}

The hole-excited states
$a_{p \uparrow} | \Psi_{\mathrm{gs}} \rangle$ and
$a_{d \uparrow} | \Psi_{\mathrm{gs}} \rangle$
belong to the two-dimensional subspace for $n_e = 1$ and $S_z = -1/2,$
which is spanned by
$
| 0; - \uparrow \rangle
\equiv
a_{p \downarrow}^\dagger
| \mathrm{vac} \rangle
$
and
$
| 1; - \uparrow \rangle
\equiv
a_{d \downarrow}^\dagger
| \mathrm{vac} \rangle
.
$
The Hamiltonian matrix for this subspace is expressed as
\begin{align}
    \mathcal{H}^{(- \uparrow)}
    =
        \begin{pmatrix}
            \varepsilon_p - \Delta \mu & t_{p d} \\
            t_{p d} & \varepsilon_d - \Delta \mu
        \end{pmatrix}
        .
    \label{Hamiltonian_down_created}
\end{align}
We map the two electronic states 
$| 0; - \uparrow \rangle$ and $| 1; - \uparrow \rangle$ to the single-qubit states
$| 0 \rangle$ and $| 1 \rangle$, respectively, when performing QPE for the spin-up-annihilated states on a quantum computer.
This mapping thus leads to the single-qubit Hamiltonian $\mathcal{H}^{(- \uparrow)}_q.$

\subsection{QAVG for one-particle GF}

\subsubsection{One-particle GF}

The one-particle GF of the dimer in frequency domain for the spin $\sigma = \uparrow, \downarrow$ sector is written as
$
G_{\sigma \kappa \kappa'} (z)
=
G_{\sigma \kappa \kappa'}^{(e) } (z)
+
G_{\sigma \kappa \kappa'}^{(h) } (z)
\ (\kappa, \kappa' = p, d)
,
$
where $z$ is a complex frequency.
We work with the zero-temperature GF throughout the present study.
Although the explanations in what follows are restricted mainly to spin-up $(\sigma = \uparrow)$ electrons for simplicity,
the quantities and expressions for spin-down $(\sigma = \downarrow)$ electrons will be defined and derived similarly.

The electron excitation part for the spin-up sector is given in the Lehmann representation by \cite{stefanucci2013nonequilibrium}
\begin{align}
    G_{\uparrow \kappa \kappa'}^{(e) } (z)
    =
        \sum_{\lambda \in (n_e = 3, \ S_z = 1/2)}
        \frac{
            \langle \Psi_{\mathrm{gs}} | a_{\kappa \uparrow} | \Psi_{\lambda} \rangle
            \langle \Psi_{\lambda} | a_{\kappa' \uparrow}^\dagger | \Psi_{\mathrm{gs}} \rangle
        }{ z - (E_{\lambda} - E_{\mathrm{gs} }) }
        ,
    \label{def_partial_G_e}
\end{align}
where $E_{\mathrm{gs}}$ is the energy eigenvalue of the ground state and $E_{\mathrm{\lambda}}$ is that of the excited state $| \Psi_\lambda \rangle.$
The hole excitation part for the spin-up sector is given by
\begin{align}
    G_{\uparrow \kappa \kappa'}^{(h) } (z)
    =
        \sum_{\lambda \in (n_e = 1, \ S_z = -1/2)}
        \frac{
            \langle \Psi_{\mathrm{gs}} | a_{\kappa' \uparrow}^\dagger | \Psi_{\lambda} \rangle
            \langle \Psi_{\lambda} | a_{\kappa \uparrow} | \Psi_{\mathrm{gs}} \rangle
        }{ z - ( E_{\mathrm{gs}} - E_{\lambda} )}
        .
    \label{def_partial_G_h}
\end{align}

The transition amplitudes in the electron excitation to the $\lambda$th eigenstate are
$
b_{\kappa \uparrow}^{(\lambda, e)}
\equiv
\langle \Psi_{\lambda} | a_{\kappa \uparrow}^\dagger | \Psi_{\mathrm{gs}} \rangle
.
$
By defining the vectors $\boldsymbol{b}_{\kappa \uparrow}^{(e)}$ via
$(\boldsymbol{b}_{\kappa \uparrow}^{(e)})_\lambda \equiv b_{\kappa \uparrow}^{(\lambda, e)},$
we find that the completeness of the excited states leads to the sum rule
\begin{align}
    \boldsymbol{b}_{\kappa \uparrow}^{(e) *}
    \cdot
    \boldsymbol{b}_{\kappa' \uparrow}^{(e)}
    =
        \delta_{\kappa \kappa'}
        -
        \gamma_{\uparrow \kappa' \kappa}
    ,
    \label{sum_rule_of_e}
\end{align}
where $\gamma_{\uparrow}$ is the one-electron density matrix defined as
$
\gamma_{\uparrow \kappa' \kappa} \equiv \langle \Psi_{\mathrm{gs}} |
a_{\kappa' \uparrow}^\dagger
a_{\kappa \uparrow}
| \Psi_{\mathrm{gs}} \rangle
.
$
The transition amplitudes in the hole excitation, on the other hand, are
$
b_{\kappa \uparrow}^{(\lambda, h)}
\equiv
\langle \Psi_{\lambda} | a_{\kappa \uparrow} | \Psi_{\mathrm{gs}} \rangle
.
$
We also find the following sum rule for them:
\begin{align}
    \boldsymbol{b}_{\kappa' \uparrow}^{(h) *}
    \cdot
    \boldsymbol{b}_{\kappa \uparrow}^{(h)}
    =
        \gamma_{\uparrow \kappa' \kappa}
    .
    \label{sum_rule_of_h}
\end{align}

The DOS of a many-electron system for the hole and electron parts are experimentally obtained via photoelectron spectroscopy (PES) and inverse PES processes, respectively \cite{bib:4070,bib:4165}.
They are to be compared under certain assumptions with the calculated DOS,
whose $\xi$ excitation ($\xi = e, h$) part as a function of excitation energy $E$ is defined as \cite{bib:7463,bib:4473,bib:4483,bib:4516,bib:4575}
\begin{align}
    \rho^{(\xi)} (E)
    =
        -\frac{1}{\pi}
        \sum_\sigma
        \mathrm{Im} \mathrm{tr}
        G_{\sigma}^{(\xi)} (E + i \delta )
        .
    \label{def_many_electron_dos}
\end{align}
$\delta$ is an infinitesimally small real constant.

\subsubsection{Natural orbitals}

The sum rules in Eqs.~(\ref{sum_rule_of_e}) and (\ref{sum_rule_of_h}) are practically unfavorable since they are not necessarily diagonal with respect to the orbital indices.
To find physically motivated parametrization of the unknown quantities later,
we introduce here the natural orbitals (NOs) \cite{bib:4643},
which are defined as the one-electron states that diagonalize the one-electron density matrix.
To this end, we get the column vector $\boldsymbol{c}_\uparrow^{(\nu)}$ for the $\nu$th eigenstate ($\nu = 0, 1$) by solving the equation
$\gamma_\uparrow \boldsymbol{c}_\uparrow^{(\nu)} = n_{\nu \uparrow} \boldsymbol{c}_{\uparrow}^{(\nu)}.$
The eigenvalue $n_{\nu \uparrow}$ represents the occupancy of the NO.
Those vectors form a orthonormalized system since $\gamma_\uparrow$ is Hermitian.
The creation operator of an electron accomodated in the $\nu$th NO is defined as
$
\widetilde{a}_{\nu \uparrow}^\dagger
=
\sum_{\kappa = p, d}
c_{\kappa \uparrow}^{(\nu) *}
a_{\kappa \uparrow}^\dagger
,
$
while the annihilation operator for the NO is defined as
$
\widetilde{a}_{\nu \uparrow}
=
\sum_{\kappa = p, d}
c_{\kappa \uparrow}^{(\nu)}
a_{\kappa \uparrow}
.
$
They satisfy the fermionic anti-commutation relations,
$
\{ \widetilde{a}_{\nu \uparrow}, \widetilde{a}_{\nu' \uparrow}^\dagger \}
=
\delta_{\nu \nu'}
.
$
Having established the NOs, we define their transition amplitudes in the electron excitation as
\begin{align}
    \widetilde{b}_{\nu \uparrow}^{(\lambda, e)}
    &\equiv
        \frac{1}{\sqrt{1 - n_{\nu \uparrow} }}
        \langle \Psi_\lambda | \widetilde{a}_{\nu \uparrow}^\dagger | \Psi_{\mathrm{gs} } \rangle
    \label{def_transition_vec_for_NO_e}
\end{align}
and those in the hole excitation as
\begin{align}
    \widetilde{b}_{\nu \uparrow}^{(\lambda, h)}
    &\equiv
        \frac{1}{\sqrt{n_{\nu \uparrow} }}
        \langle \Psi_\lambda | \widetilde{a}_{\nu \uparrow} | \Psi_{\mathrm{gs} } \rangle
    .
    \label{def_transition_vec_for_NO_h}
\end{align}
By using Eqs.~(\ref{sum_rule_of_e}) and (\ref{sum_rule_of_h}) and taking the summation over all the possible excited states,
we find the following tractable orthonormalization conditions:
\begin{align}
    \widetilde{\boldsymbol{b} }_{\nu \uparrow}^{(e) *}
    \cdot
    \widetilde{\boldsymbol{b} }_{\nu' \uparrow}^{(e)}
    =
        \widetilde{\boldsymbol{b} }_{\nu \uparrow}^{(h) *}
        \cdot
        \widetilde{\boldsymbol{b} }_{\nu' \uparrow}^{(h)}
    =
        \delta_{\nu \nu'}
    .
    \label{sum_rule_of_trans_ampls_for_NOs}
\end{align}
We define
$
d_{\kappa \uparrow}^{(\nu, e)}
\equiv
c_{\kappa \uparrow}^{(\nu) }
\sqrt{ 1 - n_{\nu \uparrow} }
$
and
$
d_{\kappa \uparrow}^{(\nu, h)}
\equiv
c_{\kappa \uparrow}^{(\nu) *}
\sqrt{ n_{\nu \uparrow} }
.
$
We can then relate the transition amplitudes for the $p$- and $d$-type orbitals and those for the NOs as
$
    b_{\kappa \uparrow}^{(\lambda, \xi)}
    =
        \sum_{\nu}
        d_{\kappa \uparrow}^{(\nu, \xi)}
        \widetilde{b}_{\nu \uparrow}^{(\lambda, \xi)}
$
for $\xi = e, h.$
From Eqs.~(\ref{def_partial_G_e}),
we can rewrite the electron part of GF as
\begin{gather}
    G_{\uparrow \kappa \kappa'}^{(e) } (z)
    =
        \sum_{\lambda, \nu, \nu'}
            d_{\kappa \uparrow}^{(\nu, e) *}
            d_{\kappa' \uparrow}^{(\nu', e) }
        \frac{
            \widetilde{b}_{\nu \uparrow}^{(\lambda, e) *}
            \widetilde{b}_{\nu' \uparrow}^{(\lambda, e)}
        }{ z - (E_{\lambda} - E_{\mathrm{gs} }) }
        ,
    \label{GF_using_NOs_e}
\end{gather}
while, from Eqs.~(\ref{def_partial_G_h}),
we can rewrite the hole part as
\begin{align}
    G_{\uparrow \kappa \kappa'}^{(h) } (z)
    =
        \sum_{\lambda, \nu, \nu'}
        d_{\kappa'}^{(\nu', h) *}
        d_{\kappa}^{(\nu, h)}
        \frac{
            \widetilde{b}_{\nu' \uparrow}^{(\lambda, e) *}
            \widetilde{b}_{\nu \uparrow}^{(\lambda, e)}
        }{ z - ( E_{\mathrm{gs}} - E_{\lambda} )}
        .
    \label{GF_using_NOs_h}
\end{align}

Although the one-electron density matrix can be obtained from the measurements on the probabilistic-excitation circuits \cite{bib:5005},
we extract it from the results of full configuration interaction (FCI) calculations since we omit the excitation process in the present study, as stated above.

\subsubsection{QPE for multiple settings}

In the present study, we use basically the standard QFT-based QPE \cite{bib:4825, bib:4826, Nielsen_and_Chuang}.
Its protocol requires the implementation of the RTE operators
\begin{align}
    U_{\mathrm{RTE}}
    \equiv
        \exp
        \left(
            -i 2 \pi \mathcal{H}_{\mathrm{QPE}}
            \frac{t_0}{N_{\mathrm{val}}}
        \right)
    \label{avr_spectra_using_QPE:def_unitary_in_QPE}
\end{align}
controlled by $n_q^{(\mathrm{QFT})}$ ancillary qubits.
The QPE circuit for a Hamiltonian $\mathcal{H}_{\mathrm{QPE}}$ provides the information about the energy eigenvalue to which an input eigenstate belongs via repeated measurements on the ancillary qubits.
The eigenvalue is designated by the peak position on the $N_{\mathrm{val}} \equiv 2^{n_q^{(\mathrm{QFT})} }$ grid points with spacing $1/t_0$ represented by the computational bases of the ancillae within certain accuracy.
When an input state is a superposition of multiple eigenstates of $\mathcal{H}_{\mathrm{QPE}}$,
multiple peaks appear whose height reflects the relative weights of the contained eigenstates.

One can easily understand that the implementation of such QPE requires $2^{n_q^{(\mathrm{QFT})}} - 1$ applications of the $U_{\mathrm{RTE}}$ operator.
The exponential increase in the operation numbers for raising the resolution motivate us strongly to find a workaround for raising the effective resolution with keeping the operation numbers.
The QAVG approach employs multiple settings for the construction of QPE circuits to alleviate the detrimental biases, e.g., spectral leakage, that may enter if a single grid mesh is adopted.
Specifically, for a given $t_0,$ we perform QPE by using a set of energy origins $\{ E_{\mathrm{orig}}^{(s)} \}_{s = 0}^{n_{\mathrm{setting}} - 1 }$ as a vernier for
$\mathcal{H}_{\mathrm{QPE}} = \mathcal{H}^{(\pm \uparrow)}.$
For an eigenvalue $\varepsilon$ of $\mathcal{H}_{\mathrm{QPE}},$
the probability of observing $| j \rangle \ (j = 0, \dots, N_{\mathrm{val}} - 1)$
when the eigenstate is input to the QPE circuit is given by \cite{Nielsen_and_Chuang}
\begin{gather}
    \mathbb{P}^{(s)}_j (\varepsilon)
    \equiv
        \left|
            \frac{1}{N_{\mathrm{val}}}
            \sum_{j' = 0}^{N_{\mathrm{val}} - 1}
            \exp
            \frac{i 2 \pi j' ((\varepsilon - E_{\mathrm{orig}}^{(s)} )t_0 - j)}{N_{\mathrm{val}} }
        \right|^2
    ,
    \label{QPE_prob_distr}
\end{gather}
which we denote as the QPE kernel in what follows.
A peak at $| j \rangle$ in a histogram for this kernel is roughly responsible for an eigenvalue $j/t_0 + E_{\mathrm{orig}}^{(s)}$ modulo $N_{\mathrm{val}}/t_0.$

When the normalized electron-excited state
$a_{\kappa \uparrow}^\dagger | \Psi_{\mathrm{gs}} \rangle/\| a_{\kappa \uparrow}^\dagger | \Psi_{\mathrm{gs}} \rangle \| \ (\kappa = p, d)$
is input to the QPE circuit for $\mathcal{H}^{(+ \uparrow)},$
the probability for observing the $\lambda$th excited state is
\begin{align}
    \mathbb{P}_\lambda^{(\kappa \uparrow, e)}
    &=
        \frac{ | \langle \Psi_\lambda | a_{\kappa \uparrow}^\dagger | \Psi_{\mathrm{gs}} \rangle |^2}{\| a_{\kappa \uparrow}^\dagger | \Psi_{\mathrm{gs}} \rangle \|^2}
    \nonumber \\
    &=
        \sum_{\nu, \nu'}
        \frac{
            d_{\kappa \uparrow}^{(\nu, e) *}
            d_{\kappa \uparrow}^{(\nu', e) }
        }{ 1 - \gamma_{\uparrow \kappa \kappa} }
            \widetilde{b}_{\nu \uparrow}^{(\lambda, e) *}
            \widetilde{b}_{\nu' \uparrow}^{(\lambda, e)}
    ,
    \label{prob_for_obs_exc_state_e}
\end{align}
where we ignored the error coming from the Trotter decomposition for simplicity.
The probability distribution $\mathbb{P}^{(s, \kappa \uparrow, e)}$ of the results from QPE is thus given by  
\begin{align}
    \mathbb{P}^{(s, \kappa \uparrow, e)}_j
    =
        \sum_{\lambda \in (n_e = 3, S_z = 1/2)}
            \mathbb{P}_\lambda^{(\kappa \uparrow, e)}
            \mathbb{P}^{(s)}_j (E_\lambda)
    ,
    \label{exact_prob_distr_e}
\end{align}
which satisfies the sum rule $\sum_j \mathbb{P}^{(s, \kappa \uparrow, e)}_j = 1,$ as expected.

When the normalized hole-excited state
$a_{\kappa \uparrow} | \Psi_{\mathrm{gs}} \rangle/\| a_{\kappa \uparrow} | \Psi_{\mathrm{gs}} \rangle \|$
is input to the QPE circuit for $\mathcal{H}^{(- \uparrow)},$
the probability for observing the $\lambda$th excited state is
\begin{align}
    \mathbb{P}_\lambda^{(\kappa \uparrow, h)}
    &=
        \frac{ | \langle \Psi_\lambda | a_{\kappa \uparrow} | \Psi_{\mathrm{gs}} \rangle |^2}{\| a_{\kappa \uparrow} | \Psi_{\mathrm{gs}} \rangle \|^2}
    \nonumber \\
    &=
        \sum_{\nu, \nu'}
        \frac{
            d_{\kappa \uparrow}^{(\nu, h) *}
            d_{\kappa \uparrow}^{(\nu', h)}
        }{ \gamma_{\uparrow \kappa \kappa} }
            \widetilde{b}_{\nu \uparrow}^{(\lambda, h) *}
            \widetilde{b}_{\nu' \uparrow}^{(\lambda, h)}
    .
    \label{prob_for_obs_exc_state_h}
\end{align}
The probability distribution $\mathbb{P}^{(s, \kappa \uparrow, h)}$ of the results from QPE is thus given by  
\begin{align}
    \mathbb{P}^{(s, \kappa \uparrow, h)}_j
    =
        \sum_{\lambda \in (n_e = 1, S_z = -1/2)}
            \mathbb{P}_\lambda^{(\kappa \uparrow, h)}
            \mathbb{P}^{(s)}_j (E_\lambda)
    ,
    \label{exact_prob_distr_h}
\end{align}
which satisfies the sum rule $\sum_j \mathbb{P}^{(s, \kappa \uparrow, h)}_j = 1.$

\subsubsection{Trial parameters for QAVG}

Here, we describe the QAVG approach for finding the one-particle GF of the dimer system via continuous parametrization.

For each combination of $s$ and $\kappa,$ a sufficiently large number of measurements should give histograms that coincide satisfactorily with the exact probability distributions $\mathbb{P}^{(s, \kappa \uparrow, e)}$ and $\mathbb{P}^{(s, \kappa \uparrow, h)}$ in Eqs.~(\ref{exact_prob_distr_e}) and (\ref{exact_prob_distr_h}), respectively.
Let $\{ f^{(s, \kappa \uparrow, \xi)}_j \}_j \ (\xi = e, h)$ be the histogram obtained from actual measurements and normalized as
$\sum_{j = 0}^{N_{\mathrm{val}} - 1} f^{(s, \kappa \uparrow, \xi)}_j = 1.$
By collecting such histograms for all the combinations of the vernier value, orbital, and excitation,
we will have the data $\{ f^{(s, \kappa \uparrow, \xi)}_j \}_{s, \kappa, \xi, j}$ based on which we estimate the many-electron DOS of the dimer model.

We assume that we already know the energy of the ground state.
The unknown quantities for estimating the DOS are then the energy eigenvalues of the excited states and the transition amplitudes.
We therefore introduce real trial parameters $\{ v_{\nu}^{(\lambda, \xi)} \}_{\nu, \lambda, \xi}$ for representing $\{ \widetilde{b}_{\nu \uparrow}^{(\lambda, \xi)} \}_{\nu, \lambda, \xi}.$
Recalling the fact that the transition amplitudes for the NOs satisfy the orthonormalization conditions in Eq.~(\ref{sum_rule_of_trans_ampls_for_NOs}) thankfully,
we parametrize them by using angles $\vartheta^{(\xi)}$ as
\begin{align}
    v_0^{(0, \xi)}
    =
        \cos \vartheta^{(\xi)}
    , \
    v_0^{(1, \xi)}
    =
        \sin \vartheta^{(\xi)}
    ,
    \nonumber \\
    v_1^{(0, \xi)}
    =
        -\sin \vartheta^{(\xi)}
    , \
    v_1^{(1, \xi)}
    =
        \cos \vartheta^{(\xi)}
    .
    \label{param_v_using_vartheta}
\end{align}
We can interpret that the two-dimensional vectors $\boldsymbol{v}_0^{(\xi)}$ and $\boldsymbol{v}_1^{(\xi)}$ can be varied by keeping the orthonormality.
By replacing the unknown amplitudes in Eqs.~(\ref{prob_for_obs_exc_state_e}) and (\ref{prob_for_obs_exc_state_h}) with the trial parameters $\{ v_{\nu}^{(\lambda, \xi)} \}_{\nu, \lambda, \xi},$
we define a function $\mathbb{P}_\lambda^{(\kappa \uparrow, \xi)} (\vartheta^{(\xi)})$ of $\vartheta^{(\xi)},$
whose explicit expressions are provided in Appendix \ref{sec:expr_for_excitation_probs}.
As for the unknown energy eigenvalues, we introduce trial parameters $\{ \varepsilon_\lambda^{(\xi)} \}_{\lambda, \xi}$ that enter the kernel in Eq.~(\ref{QPE_prob_distr}).
The theoretical probability distributions of QPE specified by the trial parameters are thus given by
\begin{align}
    \mathbb{P}^{(s, \kappa \uparrow, \xi)}_j
    (\Lambda^{(\xi)})
    \equiv
        \sum_\lambda
        \mathbb{P}_\lambda^{(\kappa \uparrow, \xi)} (\vartheta^{(\xi)})
        \mathbb{P}^{(s)}_j (\varepsilon^{(\xi)}_\lambda)
    ,
    \label{prob_distr_for_traial_params}
\end{align}
where
$
\Lambda^{(\xi)}
\equiv
\{ \vartheta^{(\xi)}, \varepsilon^{(\xi)}_0, \varepsilon^{(\xi)}_1 \}
$
represents the trial parameters collectively.
We can reconstruct the GF from Eqs.~(\ref{GF_using_NOs_e}) and (\ref{GF_using_NOs_h}) for given values of the trial parameters.

We define the cost functions in the present study based on the degree to which the observed histograms coincide with the theoretical probability distributions for the trial parameters.
Specifically, we define the dimensionless cost function for the $\xi$ excitations by averaging the contributions over the vernier as
\begin{gather}
    F_\xi (\Lambda^{(\xi)})
    \nonumber \\
    \equiv
        \frac{1}{2 n_{\mathrm{setting}} }
        \sum_s
        \sum_{\kappa = p, d}
        D
        \left(
            \mathbb{P}^{(s, \kappa \uparrow, \xi)} (\Lambda^{(\xi)})
            ,
            f^{(s, \kappa \uparrow, \xi)}
        \right)
        ,
    \label{def_cost_func}
\end{gather}
where some functional $D$ measures the discrepancy between two distributions.
The cost function has period of $N_{\mathrm{val}}/t_0$ for each of $\varepsilon^{(\xi)}_0$ and $\varepsilon^{(\xi)}_1.$
In the present study,
we adopt the $L_1$ distance \cite{Nielsen_and_Chuang}
$
D (h, h')
\equiv
\sum_{j = 0}^{N_{\mathrm{val}} - 1}
| h_j - h'_j |/2
$
between two given distributions $h$ and $h'.$
The value of $D$ falls between $0$ and $1,$
and so does the cost.
The QAVG approach finds the optimal values of the three trial parameters by minimizing the cost:
\begin{align}
    \Lambda_{\mathrm{opt}}^{(\xi)}
    =
        \argmin F_\xi (\Lambda^{(\xi)}) 
        .
\end{align}
Differentiable functional forms with respect to the trial parameters are also possible for $D,$ such as the classical infidelity \cite{Nielsen_and_Chuang} and the negative (log-)likelihood.
Gradient-based minimization algorithms can be employed for such cases.

\subsection{Physical and logical circuits}

\subsubsection{Steane code}

\begin{figure*}
\begin{center}
\includegraphics[width=17cm]{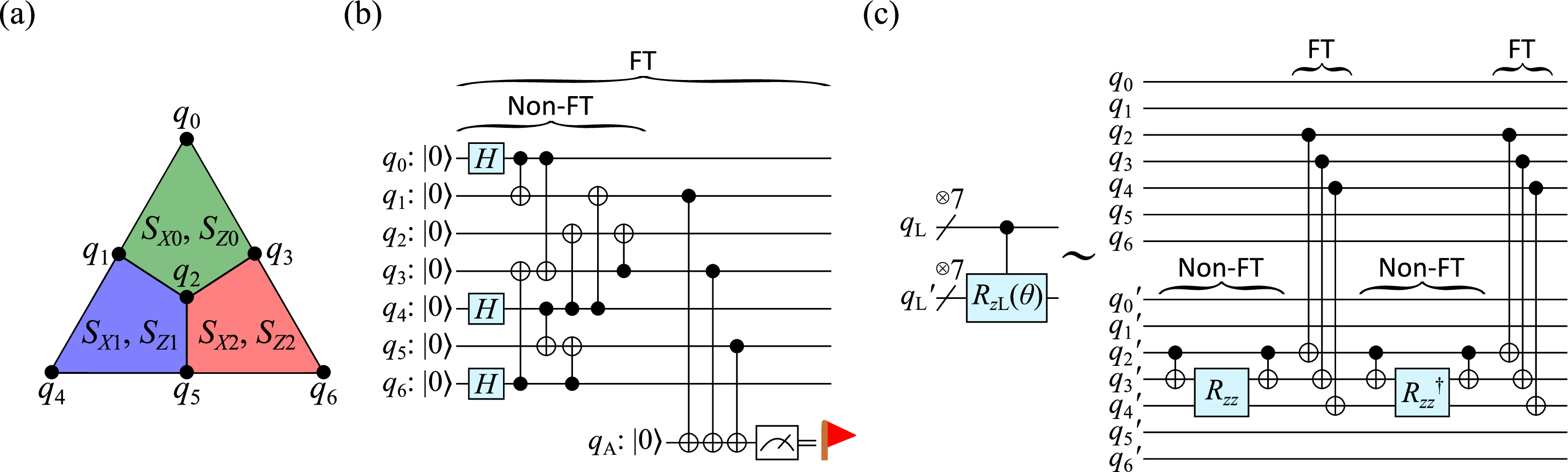}
\end{center}
\caption{
(a)
Convention for the Steane code adopted in the present study.
The seven vertices represent the constituent physical qubits forming a logical qubit.
The four vertices forming each plaquette of a single color designate the stabilizer generator for each of $X$- and $Z$-types [see Eq.~(\ref{stab_gen_of_Steane})].
This figure does not necessarily represent the actual layout of qubits on hardware.
(b)
The FT encoding circuit $U_{\mathrm{enc}}$ \cite{bib:6459} for preparing $| 0 \rangle_{\mathrm{L}}$ from initialized physical qubits.
Its former part, that encodes $| 0 \rangle_{\mathrm{L}},$ is non-FT.
The latter part prevents a harmful error from propagating outward the eight-qubit circuit by detecting it via the ancillary state $| 1 \rangle$ as a flag. 
The entire circuit is FT.
(c)
The stabilizer equivalence for a logical $z$ rotation $R_{z \mathrm{L}} (\theta)$ on a logical qubit $q_{\mathrm{L}}'$ controlled by $q_{\mathrm{L}}$ in the left circuit allows us to implement it as the right circuit. 
$R_{zz}$ in the figure represents a physical twofold rotation $R_{zz} (\theta/2) \equiv \exp(-i (\theta/4) Z \otimes Z).$
This implementation is non-FT.
}
\label{fig:Steane_code_FT_enc}
\end{figure*}

The Steane code \cite{bib:5281} is a $[[7,1,3]]$ Calderbank--Shor--Steane (CSS) code \cite{bib:6432}, capable of correcting a single-qubit error.
It is also known as the smallest one belonging to topological two-dimensional color codes \cite{bib:6370}.
We adopt the following six Pauli operators as the stabilizer generators for this code:
\begin{align}
\begin{split}
    S_{X0}
    &=
        X_0 X_1 X_2 X_3,
    \\
    S_{X1}
    &=
        X_1 X_2 X_4 X_5,
    \\
    S_{X2}
    &=
        X_2 X_3 X_5 X_6,
    \\
    S_{Z0}
    &=
        Z_0 Z_1 Z_2 Z_3,
    \\
    S_{Z1}
    &=
        Z_1 Z_2 Z_4 Z_5,
    \\
    S_{Z2}
    &=
        Z_2 Z_3 Z_5 Z_6
        ,
\end{split}
    \label{stab_gen_of_Steane}
\end{align}
as depicted in Fig.~\ref{fig:Steane_code_FT_enc}(a).
These operators define the code space where possible states of a single logical qubit reside.
That is, the necessary and sufficient condition for a physical seven-qubit state to be a logical single-qubit state is that this state is a simultaneous eigenstate belonging to the subspace with $+1$ eigenvalues of the six stabilizer generators. 
Logical $X$ and $Z$ gates can be defined as the tensor products of physical gates as
$X_{\mathrm{L}} \equiv X^{\otimes 7}$ and $Z_{\mathrm{L}} \equiv Z^{\otimes 7},$ respectively.
It is known that logical $H, S, \mathrm{CNOT}$ gates for this code,
which enable us to implement any logical Clifford operation, admit transversal implementation: $H_{\mathrm{L}} \equiv H^{\otimes 7}, S_{\mathrm{L}} \equiv S^{\dagger \otimes 7},$ and $\mathrm{CNOT}_{\mathrm{L}} \equiv \mathrm{CNOT}^{\otimes 7}.$
This fact tells us that non-transversal implementation of logical single-qubit non-Clifford operations, e.g., $T$ gates and logical $z$ rotation gates for generic angles, inevitably appear for any logical universal gate set due to the Eastin--Knill theorem \cite{bib:7036}.
Since non-transversal implementation on a circuit can render the circuit non-FT,
it is desirable to find a way for performing non-Clifford operations in an FT manner.
There exist two well known approaches for that.
The first one is the recursive gate teleportation \cite{bib:7292, bib:7293, bib:7294, bib:6542} starting from state injection.
The second one is code switching \cite{bib:7451, bib:7321}, which employs another code capable of FT non-Clifford operations.
These approaches, however, incur large resource overheads.
We therefore decide in the present study to adopt the direct implementation of a logical $z$ rotation as
$R_{z \mathrm{L}} (\theta) \equiv \exp(- i (\theta/2) Z_{\mathrm{L}})$
for a rotation angle $\theta,$
which is non-FT.

For the preparation of a logical state $| 0 \rangle_{\mathrm{L}},$
we adopt the FT encoding circuit $U_{\mathrm{enc}}$ proposed by Goto \cite{bib:6459},
which uses a single physical flag qubit for detecting harmful errors that may occur on the eight-qubit circuit involving 11 physical CNOT gates,
as shown in Fig.~\ref{fig:Steane_code_FT_enc}(b).

We can reduce the weight of the exponent by using the stabilizer equivalence (denoted by a symbol $\sim$) for $Z_{\mathrm{L}}.$
For example, we have $Z_{\mathrm{L}} \sim Z_2 Z_3 Z_4$ since $S_{Z0} S_{Z2} Z_{\mathrm{L}} = Z_2 Z_3 Z_4,$ which means
$R_{z \mathrm{L}} (\theta) \sim \exp(- i (\theta/2) Z_2 Z_3 Z_4).$
We can thus replace the direct implementation of the logical rotation by the weight-3 operator, which is still non-FT.
The stabilizer equivalence can also be used for implementing a logical controlled rotation.
For example, a logical controlled $z$ rotation can be implemented as shown in Fig.~\ref{fig:Steane_code_FT_enc}(c),
where each $\mathrm{CNOT}_{\mathrm{L}}$ requires only three physical CNOTs \cite{bib:7425}.

\subsubsection{Physical circuits for QPE sampling}

\begin{figure*}
\begin{center}
\includegraphics[width=16cm]{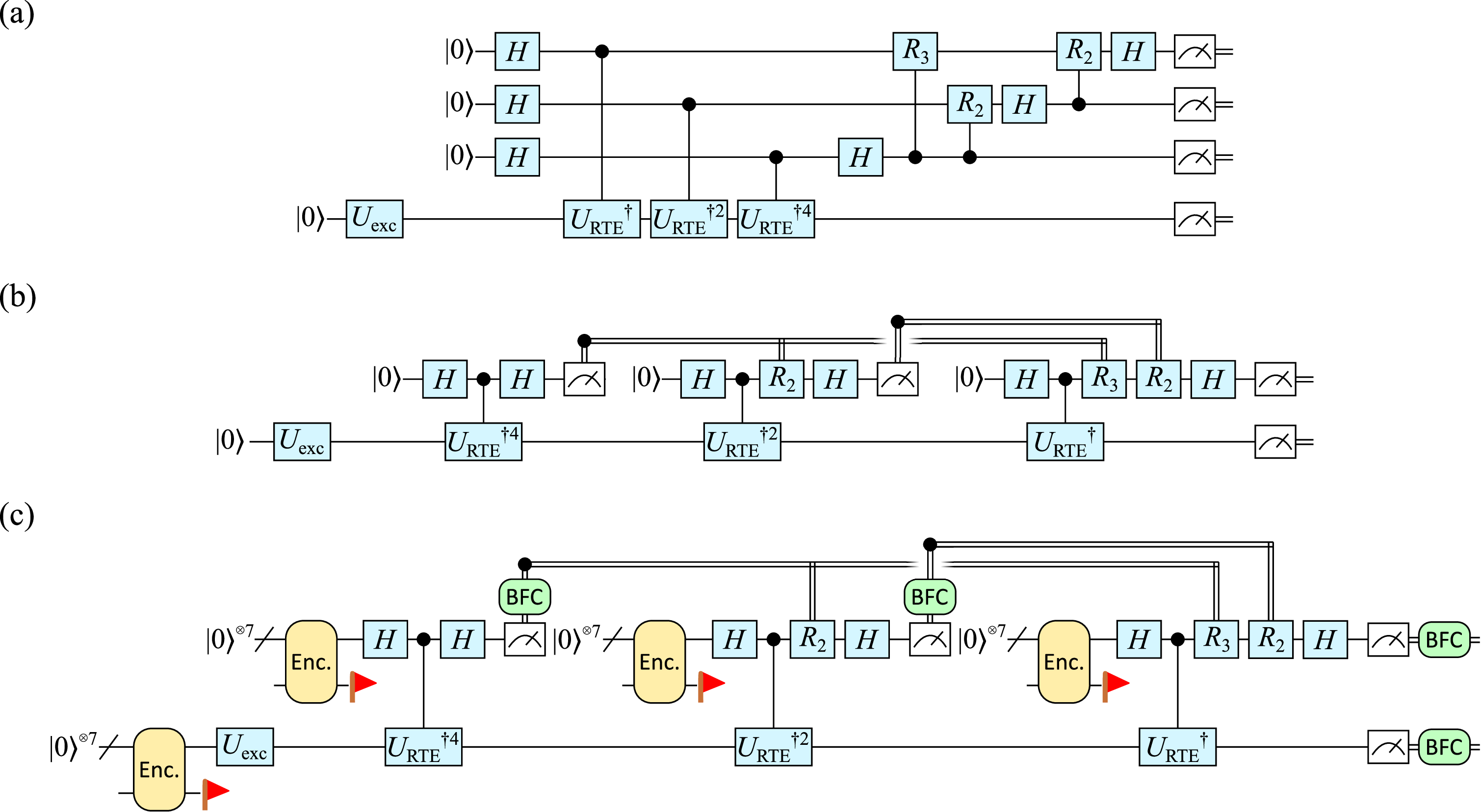}
\end{center}
\caption{
(a)
Four-qubit physical circuit $\mathcal{C}^{(\mathrm{phys})}$ for the QPE sampling used in the present study.
$U_{\mathrm{exc}}$ prepares the electron- or hole-excited state.
This performs QPE for $n_q^{(\mathrm{QFT})} = 3$ by using three ancillary qubits with $R_k \equiv \mathrm{diag} (1, \exp(-i 2 \pi/2^k)).$
The phase gates to adjust the origin $E_{\mathrm{orig}}^{(s)}$ of energy for the $s$th setting are not drawn in the figure for simplicity.
(b)
Two-qubit physical circuit $\mathcal{C}^{(\mathrm{phys,1a})}.$
This performs QPE sampling effectively for $n_q^{(\mathrm{QFT})} = 3$ by using a single ancillary qubit via classically controlled operations.
(c)
Logical circuit $\mathcal{C}^{(\mathrm{log,1a})}$ for the QPE sampling.
The gates of the same names as in the physical circuits are actually the logical versions of the corresponding physical gates.
The encoding of a $| 0 \rangle_{\mathrm{L}}$ state is performed by using the $U_{\mathrm{enc}}$ circuit in Fig.~\ref{fig:Steane_code_FT_enc}(b) and measuring a flag qubit.
}
\label{fig:qpe_circuits}
\end{figure*}

The most naive physical circuit $\mathcal{C}^{(\mathrm{phys})}$ for QPE sampling in the present study consists of four qubits, as shown in Fig.~\ref{fig:qpe_circuits}(a).
It begins with the excited-state preparation $U_{\mathrm{exc}}$ for $a_{\kappa \uparrow}^\dagger | \Psi_{\mathrm{gs}} \rangle$ or $a_{\kappa \uparrow} | \Psi_{\mathrm{gs}} \rangle \ (\kappa = p, d)$ mapped to a single qubit.
$U_{\mathrm{exc}}$ is actually a $y$ rotation whose angle is provided in Appendix \ref{sec:FCI_results}.
The excited state is then fed into the (inverse) QPE part,
which is the ordinary QFT-based one \cite{Nielsen_and_Chuang},
The computational bases of $n_q^{(\mathrm{QFT})} = 3$ ancillae represent directly the grid points for estimating the eigenvalues of $\mathcal{H}^{\mathrm{QPE}}.$
We can adjust the energy origin $E_{\mathrm{orig}}^{(s)}$ by introducing phase gates to the ancillae.
It is the case for all the QPE circuits in the present study.

The ancilla number can be reduced by performing the measurements sequentially and reusing a single ancilla.
Specifically, we can also perform the QPE sampling by using the two-qubit circuit $\mathcal{C}^{(\mathrm{phys,1a})}$ shown in Fig.~\ref{fig:qpe_circuits}(b),
which employs midcircuit measurements for classically controlled operations \cite{bib:7059,bib:7325,bib:7060,bib:7441}.

\subsubsection{Logical circuits for QPE sampling}

The logical version $\mathcal{C}^{(\mathrm{log,1a})}$ of the single-ancilla physical circuit $\mathcal{C}^{(\mathrm{phys,1a})}$ is shown in Fig.~\ref{fig:qpe_circuits}(c).
There exist two unique features in the logical circuit.
The first one is the incorporation of the FT encoding by employing $U_{\mathrm{enc}}$ with flag qubits in Fig.~\ref{fig:Steane_code_FT_enc}(b).
The second one is the offline bit-flip correction (BFC) for observed classical bit strings.
To be specific, we assume that seven classical bits observed in a $Z_{\mathrm{L}}$ measurement on a logical register may have suffered from at most a single flip.
Such a flip can originate from a flip of a constituent physical qubit somewhere before the measurement and from a readout error.
Regardless of the kind of the origin, we can know whether a single flip has occurred or not and identify which classical bit has been flipped.
We can thus correct the error by making a lookup table before the circuit execution for possible observed seven classical bits by referring to the 16 computational bases contained in $| 0 \rangle_{\mathrm{L}}$ or $| 1 \rangle_{\mathrm{L}}.$

\section{Results and discussion}

\subsection{Computational details}

\subsubsection{Classical computation}

We first performed DFT-based geometry optimization for a system consisting of a $\chi$-Fe$_5$C$_2$ slab with a CO molecule adsorbed onto a hollow site of the surface.
Specifically, we adopted the unit cell whose basal plane with edge lengths $a = 5.10$ and $b = 12.88$ \AA, forming an angle of $86.3^\circ.$
We set the edge length $c$ of the unit cell perpendicular to the basal plane to $12.73$ \AA,
so that the width of the vacuum region was $3.5$ \AA.
The unit cell contained 40 Fe atoms and 16 C atoms hewed from the bulk.
The positions of the 58 atoms (including the CO molecule) were relaxed by using VASP \cite{bib:VASP} based on the projector augmented wave (PAW) method for spin-unpolarized DFT calculations with the Perdew--Burke--Ernzerhof (PBE) \cite{bib:37} functional.

After obtaining the optimized geometry, we performed a self-consistent calculation for $4 \times 1 \times 1$ $k$ points by using Quantum Espresso \cite{Quantum_Espresso_refs_1,Quantum_Espresso_refs_2,Quantum_Espresso_refs_3} with norm-conserving pseudopotentials \cite{bib:4555}.
We set the cutoff energy of plane waves to $150$ eV for expanding the Bloch wave functions.
We constructed the MLWOs by using RESPACK \cite{bib:7438}.
The screened Coulomb repulsion between the MLWOs were calculated based on the cRPA approach by using the dielectric function calculated from 500 valence bands.
We set the cutoff energy for the dielectric function to 4 Ry.
We confirmed that the calculated screened Coulomb repulsion energies are almost the same even when 1000 valence bands are taken into account.

\subsubsection{Quantum computation}

We constructed the circuits that may contain midcircuit measurements and control-flow operations by using Qiskit \cite{qiskit} and converted them into the format for pytket \cite{bib:7467}.
We used Quantinuum H2-2, a trapped-ion quantum computer \cite{bib:7169,bib:7466}, for the experiments.
By making use of the Quantinuum Nexus platform \cite{quantinuum_nexus},
we transpiled the circuit data for the native gate set of the hardware and then submitted them for execution.

\subsection{DFT calculations for construction of MLWOs}

By examining the projected DOS and the Bloch wave functions,
we found that the bottom of the valence bands involves the significant contributions from the $5 \sigma$ and $1 \pi$ MOs,
hybridized with the delocalized states in Fe$_5$C$_2$ [see Fig.~\ref{fig:Fe5C2-CO_cell_and_dft_dos}(c)]. 
While the $5 \sigma$ orbital in an isolated CO molecule is higher in energy than the $1 \pi$ orbitals,
it is lower in the adsorption system.

\begin{figure*}
\begin{center}
\includegraphics[width=14cm]{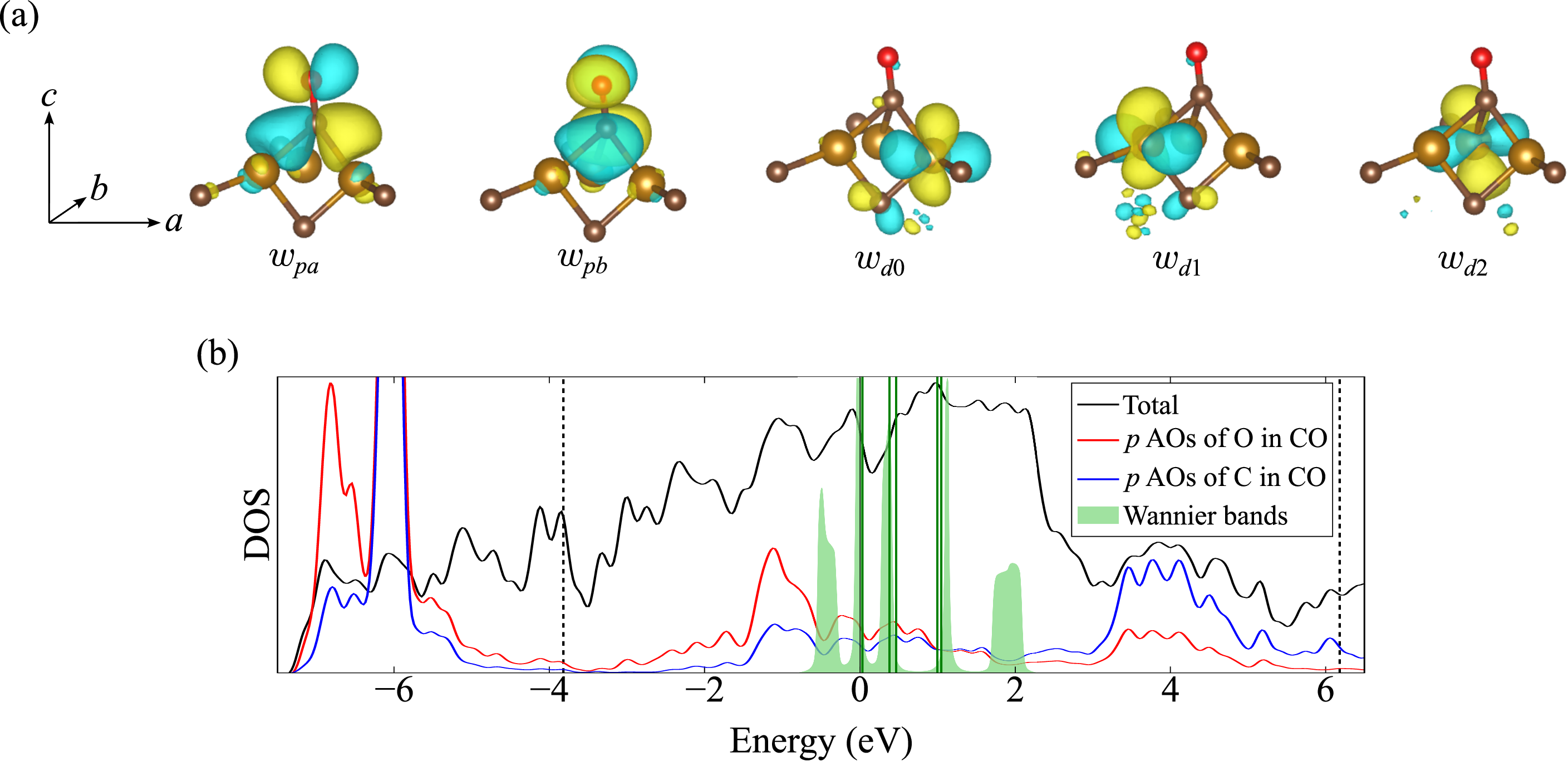}
\end{center}
\caption{
(a)
Five MLWOs constructed from the DFT calculation.
(b)
Electronic DOS originating from the interpolated Wannier bands are shown as shaded region.
The total DOS and projected DOS in this figure are the same as in 
Fig.~\ref{fig:Fe5C2-CO_cell_and_dft_dos}.
The energy range between the two vertical dashed lines is the energy window for the construction of the MLWOs.
The vertical green solid lines indicate the orbital energies of the MLWOs.
}
\label{fig:Fe5C2-CO_dos_wannier}
\end{figure*}

We defined the energy window for constructing MLWOs as covering 10 eV range around the Fermi level, so that the Bloch wave functions were incorporated from the range.
When minimizing the total spread of trial Wannier orbitals,
we started from an initial guess that consisted of $p_x$- and $p_y$-shaped orbitals (parallel to the basal plane) around the C atom in the CO molecule and three spherical amplitudes in the middle of Fe-C bonds linking the molecule and surface.
The resultant MLWOs are drawn in Fig.~\ref{fig:Fe5C2-CO_dos_wannier}(a).
As shown in the figure, we obtained the CO $2 \pi^*$-derived MLWOs, denoted as $w_{p a}$ and $w_{p b},$
and the Fe $d$-derived MLWOs, denoted as $w_{d 0}, w_{d 1},$ and $w_{d 2}.$
$w_{p a}$ ($w_{p b}$) is basically of anti-bonding nature of the $p$ orbitals of C and O atoms along the $a$ ($b$) direction.
Although the shapes of $w_{p a}$ and $w_{p b}$ resemble each other, they actually differ due to the asymmetric local geometry.
Similarly, the shapes of $w_{d 0}, w_{d 1},$ and $w_{d 2}$ are slightly different from each other.
The parameters of the five MLWOs are shown in Table \ref{tab:params_of_wannier}.
It is found that the screened Coulomb repulsion energies are much smaller than the bare values,
indicating the necessity of incorporation of screening effects for constructing a realistic model.
In addition, even the values of screened repulsion are much larger than the transfer integrals.
This fact implies that the electronic correlation is crucial in this adsorption system.
The DOS for the Wannier-interpolated bands are shown in Fig.~\ref{fig:Fe5C2-CO_dos_wannier}(b),
 where we see that the orbital energies are within a range of $\approx 1$ eV despite the wide energy window.

As mentioned above, we considered a two-orbital model consisting of $p$- and $d$-type orbitals.
We defined the parameters of this model for $\mathcal{H}^{(\mathrm{dimer})}$ based on the average values of the five MLWOs and their screened repulsion energies.
We set the shift $\Delta \mu$ of the chemical potential to $1.5$ eV in what follows,
for which we confirmed that the ground state is a two-electron state.
The parameters of the dimer Hamiltonian and those of the corresponding single-qubit Hamiltonians $\mathcal{H}^{(\pm \uparrow)}_q$ coming from Eqs.~(\ref{Hamiltonian_up_created}) and (\ref{Hamiltonian_down_created}) are shown in Table \ref{tab:hamiltonian_params}.

\begin{table*}[]
\centering
\caption{Orbital energies, transfer integrals, and on-site Coulomb repulsion energies of the five MLWOs centered at the same unit cell. The orbital energies are those measured from the Fermi level found in the DFT calculation. The tabulated values are all in eV. }
\label{tab:params_of_wannier}
\begin{tabular}{cclccccclcc}
\hline
      & Orbital energy &  & \multicolumn{5}{c}{Transfer integral}                &  & \multicolumn{2}{c}{On-site repulsion} \\ \cline{4-8} \cline{10-11} 
      &                &  & $p a$    & $p b$    & $d 0$    & $d 1$    & $d 2$    &  & Bare               & cRPA             \\ \hline
$p a$ & $0.996$        &  &          & $0.114$  & $-0.735$ & $0.318$  & $0.037$  &  & $12.05$            & $1.94$           \\
$p b$ & $1.047$        &  & $0.114$  &          & $-0.373$ & $-0.185$ & $-0.235$ &  & $12.06$            & $1.98$           \\
$d 0$ & $0.464$        &  & $-0.735$ & $-0.373$ &          & $-0.647$ & $0.021$  &  & $18.54$            & $2.26$           \\
$d 1$ & $0.380$        &  & $0.318$  & $-0.185$ & $-0.647$ &          & $-0.029$ &  & $18.29$            & $2.17$           \\
$d 2$ & $0.031$        &  & $0.037$  & $-0.235$ & $0.021$  & $-0.029$ &          &  & $18.69$            & $2.23$           \\ \hline
\end{tabular}
\end{table*}

\begin{table*}[]
\centering
\caption{Parameters of the dimer Hamiltonian $\mathcal{H}^{(\mathrm{dimer})}$ extracted from the MLWOs and those of the single-qubit Hamiltonians $\mathcal{H}^{(+ \uparrow)}_q$ and $\mathcal{H}^{(- \downarrow)}_q$ for quantum computation. The orbital energies are those measured from the Fermi level found in the DFT calculation. The tabulated values are all in eV.}
\label{tab:hamiltonian_params}
\begin{tabular}{ccccccccccccc}
\hline
\multicolumn{5}{c}{$\mathcal{H}^{(\mathrm{dimer})}$}           &  & \multicolumn{3}{c}{$\mathcal{H}^{(+ \uparrow)}_q$} &  & \multicolumn{3}{c}{$\mathcal{H}^{(- \uparrow)}_q$} \\ \cline{1-5} \cline{7-9} \cline{11-13} 
$\varepsilon_p$ & $\varepsilon_d$ & $t_{pd}$ & $U_p$  & $U_d$  &  & $h_0$                     & $h_x$      & $h_z$     &  & $h_0$                   & $h_x$       & $h_z$      \\ \hline
$1.021$         & $0.292$         & $-0.195$ & $1.96$ & $2.22$ &  & $4.060 - 3 \Delta \mu$    & $-0.195$   & $0.234$   &  & $0.657 - \Delta \mu$    & $-0.195$    & $0.365$    \\ \hline
\end{tabular}
\end{table*}

\subsection{QAVG from physical QPE}

\subsubsection{Circuit execution}

\begin{figure*}
\begin{center}
\includegraphics[width=17cm]{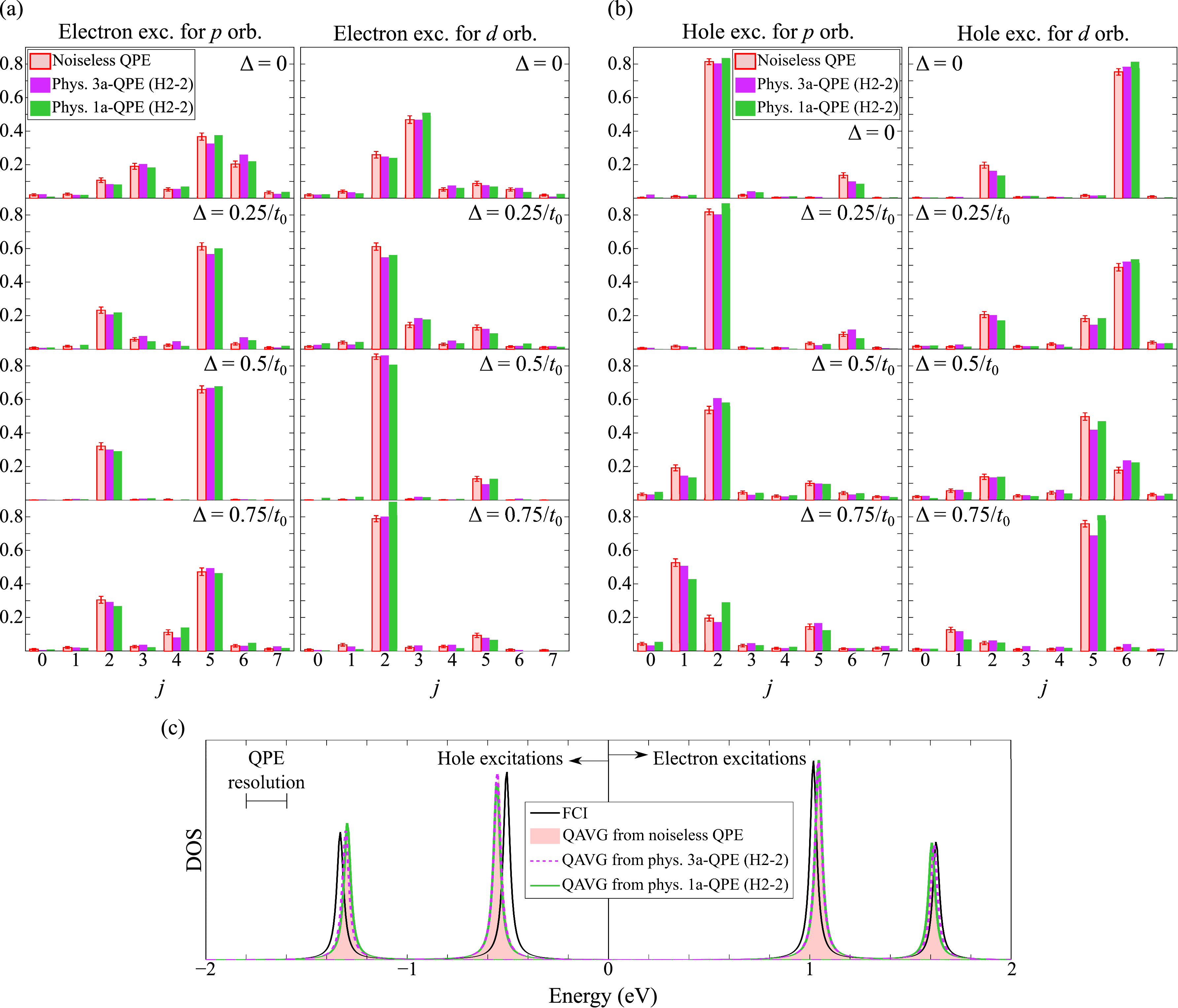}
\end{center}
\caption{
(a)
Normalized histograms obtained from the three-ancilla QPE circuit $\mathcal{C}^{\mathrm{(phys)}}$ and the single-ancilla QPE circuit $\mathcal{C}^{\mathrm{(phys, 1a)}}$ for electron excitations with four origin shifts $\Delta$ executed on H2-2.
For each of the histograms, $n_{\mathrm{shots}} = 500$ shots were performed.
$j$ represents each measurement outcome indicating the observed (effective) three ancillae.
The probability distributions $\mathbb{P}_j$ calculated from noiseless circuit simulations are also shown,
together with the standard deviations $\sigma_j = \sqrt{\mathbb{P}_j (1 - \mathbb{P}_j)/n_{\mathrm{shots}}}$ as error bars.
(b)
Histograms and related quantities obtained for hole excitations.
(c)
Many-electron DOS obtained from the QAVG calculations using the histograms with the four shifts.
Those from the FCI results and the noiseless circuit simulations are also shown.
The QPE resolution as the grid spacing $1/t_0$ is depicted in the figure.
}
\label{fig:qavg_fci_and_phys_qpe}
\end{figure*}

We first collected histograms from physical QPE on H2-2.
Specifically, we executed the three-ancilla (3a) QPE circuits $\mathcal{C}^{\mathrm{(phys)}}$
in Fig.~\ref{fig:qpe_circuits}(a)
and the single-ancilla (1a) QPE circuits $\mathcal{C}^{\mathrm{(phys, 1a)}}$
in Fig.~\ref{fig:qpe_circuits}(b).
We set the energy origin for the QPE circuits as
$E_{\mathrm{orig}}^{(s)} \equiv E_{\mathrm{o}} + \Delta$ with $E_{\mathrm{o}} = -0.8$ eV and grid spacing $1/t_0 = 0.2$ eV,
where $\Delta \equiv s/(4 t_0) \ (s = 0, \dots, 3)$ is the origin shift.
For each of the histograms, we accumulated the measurement outcomes for $n_{\mathrm{shots}} = 500$ shots.
Figure \ref{fig:qavg_fci_and_phys_qpe}(a) and (b) show the normalized histograms
$\{ f^{(s, \kappa, \xi)}_j \}_{s, \kappa, j}$ for the electron $(\xi = e)$ and hole $(\xi = h)$ excitations, respectively.
The probability distributions calculated from state vectors based on noiseless circuit simulations are also shown.
It is seen that the histograms collected from the real quantum computer are in good agreement with the noiseless probability distributions.
In addition, the discrepancy between the 3a-QPE results and the noiseless results does not much differ from that between the 1a-QPE and the noiseless results.

\subsubsection{QAVG calculations}

Having collected the histograms, we minimized the cost values defined in Eq.~(\ref{def_cost_func}) for the electron and hole excitations separately by running the Nelder-Mead algorithm hundreds of times for randomly generated initial trial parameters.
The optimized values among possibly multiple degenerate global minima are provided in Table \ref{tab:qavg_optimized_params}.
We then reconstructed the GF from the optimized trial parameters $\Lambda^{(\xi)}_{\mathrm{opt}}$ and substituted them into Eq.~(\ref{def_many_electron_dos}) with $\delta = 0.02$ eV as a smearing constant to get the many-electron DOS,
as plotted in Fig.~\ref{fig:qavg_fci_and_phys_qpe}(c).
The DOS from the FCI results and the noiseless circuit simulations are also shown in the figure.
It is noted here that the FCI-DOS does not coincide exactly with the noiseless QAVG-DOS due to the Trotterization errors that slipped into the latter.
The noisy QAVG-DOS should hence coincide with the noiseless QAVG-DOS ideally.
Those from the 3a- and 1a-QPE are actually in very good agreement with the noiseless DOS.

\begin{figure*}
\begin{center}
\includegraphics[width=17cm]{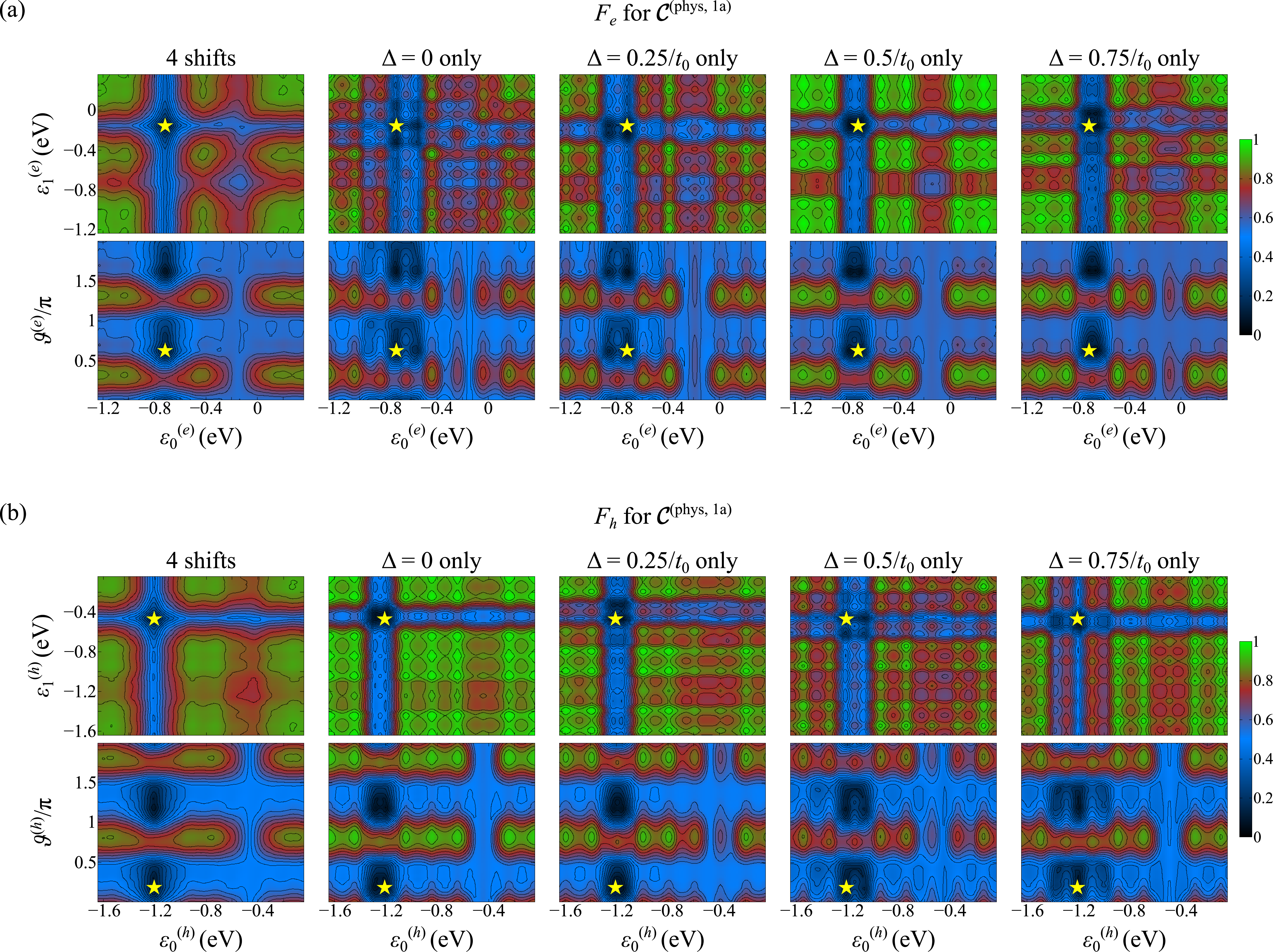}
\end{center}
\caption{
(a)
Cost landscapes for the electron excitations based on the histograms collected from the single-ancilla QPE circuits $\mathcal{C}^{\mathrm{(phys, 1a)}}$ on H2-2.
In the leftmost column,
the QAVG cost is calculated based on the four shifts $\Delta$ of energy origin in the QPE circuit
and the upper panel represents
$F_e (\vartheta^{(e)}_{\mathrm{opt}}, \varepsilon^{(e)}_0, \varepsilon^{(e)}_1),$
while the lower panel represents
$F_e (\vartheta^{(e)}, \varepsilon^{(e)}_0, \varepsilon^{(e)}_{1, \mathrm{opt}}).$
Each one of the other four columns represents the single-reference cost landscape based on only one of the four shifts.
The star symbols designate the optimized parameters $\Lambda^{(e)}_{\mathrm{opt}}$ for $\mathcal{C}^{\mathrm{(phys, 1a)}}$ in Table \ref{tab:qavg_optimized_params}.
(b) Similar plots for the hole excitations.
}
\label{fig:qavg_cost_for_phys_qpe}
\end{figure*}

The QAVG cost landscape $F_e (\vartheta^{(e)}, \varepsilon^{(e)}_0, \varepsilon^{(e)}_1)$ for the electron excitations,
which was used in the optimization by referring to the histograms for the four origin shifts,
are plotted in the leftmost column of Fig.~\ref{fig:qavg_cost_for_phys_qpe}(a).
The single-reference cost landscapes that are defined by referring to only one of the four origin shifts are also plotted in the figure.
We can see that the single-reference landscapes are more rugged than the QAVG landscape.
Furthermore, the single-reference landscapes around the location of the QAVG-optimized parameters $\Lambda^{(e)}_{\mathrm{opt}}$ suffer from multiple minima.
These features come from the oscillatory behavior, or equivalently the spectral leakage, of the QPE kernel and it is also the case for generic systems.
This fact implies that QPE sampling and subsequent cost minimization tends to be stuck in local minima more often for larger systems.
We should thus wash out the rugged features of a cost landscape as much as possible by resorting to the QAVG approach.

\begin{table*}[]
\centering
\caption{Optimized parameters $\Lambda^{(e)}_{\mathrm{opt}}$ and $\Lambda^{(h)}_{\mathrm{opt}}$ for the QAVG cost functions. The functional forms of the cost differ from each other depending on the QPE circuits for collecting the histograms. The average $L_1$ distances between the noiseless probability distributions and the observed histograms are also shown. The tabulated energy values are all in eV. }
\label{tab:qavg_optimized_params}
\begin{tabular}{cccccclcccc}
\hline
Circuit                                    & Average $L_1$ dist.  & \multicolumn{4}{c}{Electron part}                                                                                         &  & \multicolumn{4}{c}{Hole part}                                                                                             \\ \cline{3-6} \cline{8-11} 
\multicolumn{1}{l}{}                       & \multicolumn{1}{l}{} & $\vartheta^{(e)}_{\mathrm{opt}}$ & $\varepsilon^{(e)}_{0,\mathrm{opt}}$ & $\varepsilon^{(e)}_{1,\mathrm{opt}}$ & Cost     &  & $\vartheta^{(h)}_{\mathrm{opt}}$ & $\varepsilon^{(h)}_{0,\mathrm{opt}}$ & $\varepsilon^{(h)}_{1,\mathrm{opt}}$ & Cost     \\ \hline
Noiseless                                  & \multicolumn{1}{l}{} & $0.6251 \pi$                     & $-0.7256$                            & $-0.1554$                            & $0.0362$ &  & $0.1693 \pi$                     & $-1.2180$                            & $-0.4690$                            & $0.0627$ \\
$\mathcal{C}^{(\mathrm{phys})}$ on H2-2    & $0.0556$             & $0.6247 \pi$                     & $-0.7204$                            & $-0.1465$                            & $0.0590$ &  & $0.1666 \pi$                     & $-1.2101$                            & $-0.4579$                            & $0.0868$ \\
$\mathcal{C}^{(\mathrm{phys,1a})}$ on H2-2 & $0.0626$             & $0.6263 \pi$                     & $-0.7193$                            & $-0.1568$                            & $0.0559$ &  & $0.1906 \pi$                     & $-1.2089$                            & $-0.4654$                            & $0.0731$ \\
$\mathcal{C}^{(\mathrm{log,1a})}$ on H2-2  & $0.1135$             & $0.5888 \pi$                     & $-0.7304$                            & $-0.1564$                            & $0.1295$ &  & $0.1363 \pi$                     & $-1.2202$                            & $-0.4627$                            & $0.1474$ \\ \hline
\end{tabular}
\end{table*}

\subsubsection{Direct reconstruction of DOS}

The rather good agreement between the ideal spectra and the observed ones in Fig.~\ref{fig:qavg_fci_and_phys_qpe}(c) may have come from the continuous parametrization and the gentle cost landscapes, as illustrated above.
Here we examine two simple ways for direct reconstruction (DR) of DOS to compare them with the QAVG-DOS.

\begin{figure*}
\begin{center}
\includegraphics[width=13cm]{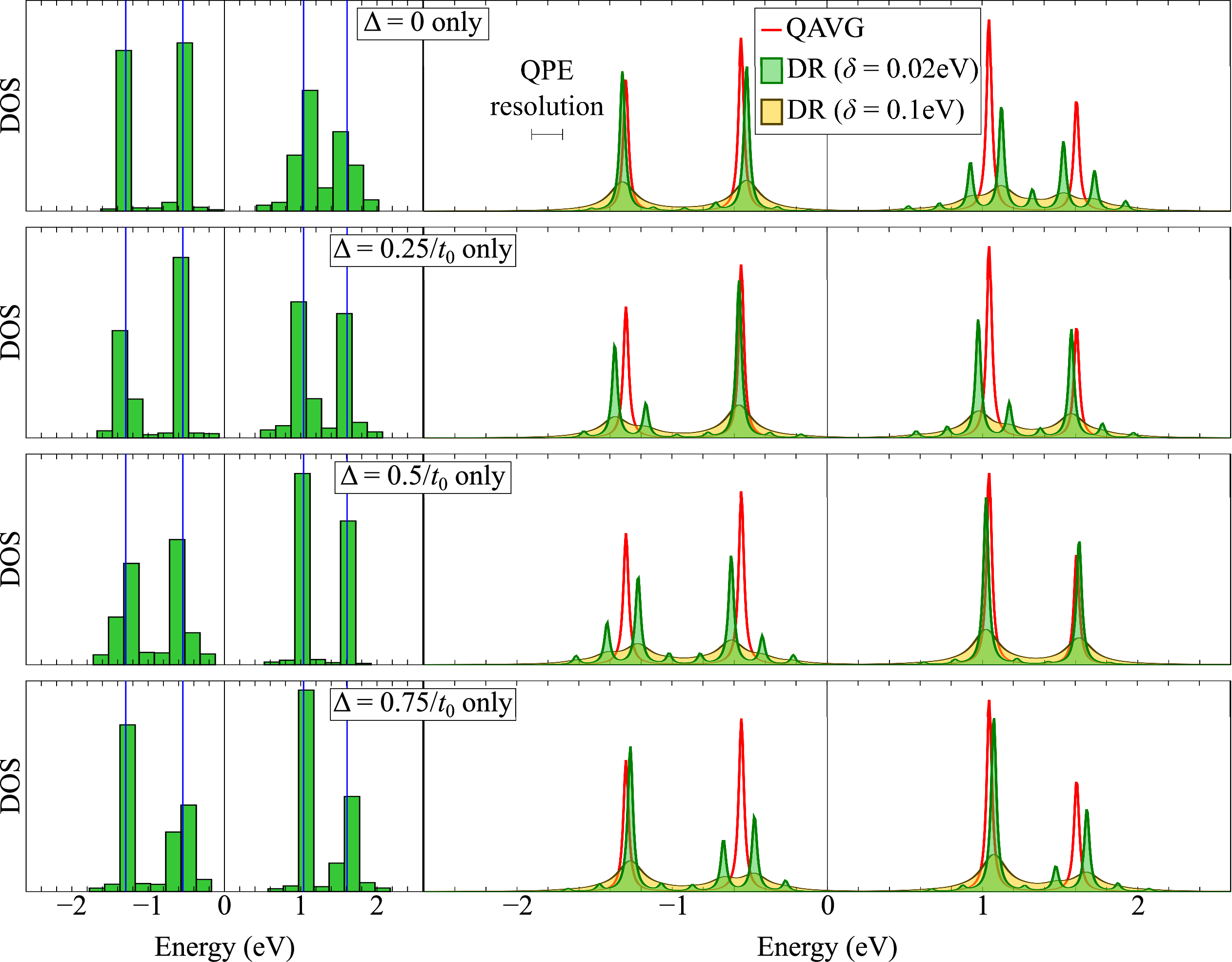}
\end{center}
\caption{
On each of the four rows,
the left part shows the direct reconstruction (DR) of DOS as bars based on the histograms from the circuits $\mathcal{C}^{(\mathrm{phys,1a})}$ on H2-2 with a single origin shift $\Delta.$
The vertical blue lines indicate the optimized parameters 
$\varepsilon^{(\xi)}_{0, \mathrm{opt}}$ and $\varepsilon^{(\xi)}_{1, \mathrm{opt}}$ for the noiseless circuits in Table \ref{tab:qavg_optimized_params}.
The right part shows the DR-DOS as sums of Lorentzians with smearing constants $\delta.$
The QAVG-DOS in Fig.~\ref{fig:qavg_fci_and_phys_qpe}(c) is also shown for comparison.
}
\label{fig:discr_reconst_dos_phys_1anc}
\end{figure*}

As the first way, we take the bars from the histograms for $\mathcal{C}^{(\mathrm{phys,1a})}$ with a single shift $\Delta$ and locate them on the continuous energy axis such that each bar distributes its frequency uniformly in the interval centered at the corresponding grid point.
We multiply the frequencies by the excitation probabilities depending on the electron- or hole-type histograms to obtain a piecewisely constant function as DR-DOS.
The results are shown in the left part of each row in
Fig.~\ref{fig:discr_reconst_dos_phys_1anc}.
While the peak locations are found to be in reasonable agreement with the optimized parameters $\varepsilon^{(\xi)}_{0, \mathrm{opt}}$ and $\varepsilon^{(\xi)}_{1, \mathrm{opt}}$ for all the origin shifts,
the relative peak intensities differ much depending on the shift.

As the second way, 
we reconstruct the DOS as a sum of Lorentzians.
Specifically, we replace the bars in the DR-DOS explained just above by functions of the form $L (E) \propto \delta/(E^2 + \delta^2)$ at the individual grid points, leading to a smooth function.
$\delta$ is a smearing constant.
This reconstruction for a large $\delta$ is expected to alleviate the biases originating from the QPE grid points and the origin shift.
The results are shown in the right part of each row in
Fig.~\ref{fig:discr_reconst_dos_phys_1anc}.
We see for the smaller smearing, $\delta = 0.02$ eV, that the strong peaks in the DR-DOS coincide with the QAVG-DOS if those peaks are occasionally (nearly) on the QPE grid points.
The DR-DOS otherwise exhibits weaker peaks around the true peak location,
which prohibits one from deciding whether the peaks come from a single true peak or not.
For the larger smearing, $\delta = 0.1$ eV, on the other hand,
the shapes of DR-DOS for all the origin shifts resemble,
exhibiting two major peaks.
They are, however, so obscure that the effective resolution of the resultant DOS has been lowered from the raw QPE results.

\subsection{QAVG from logical QPE}

\subsubsection{Circuit execution}

\begin{figure*}
\begin{center}
\includegraphics[width=12cm]{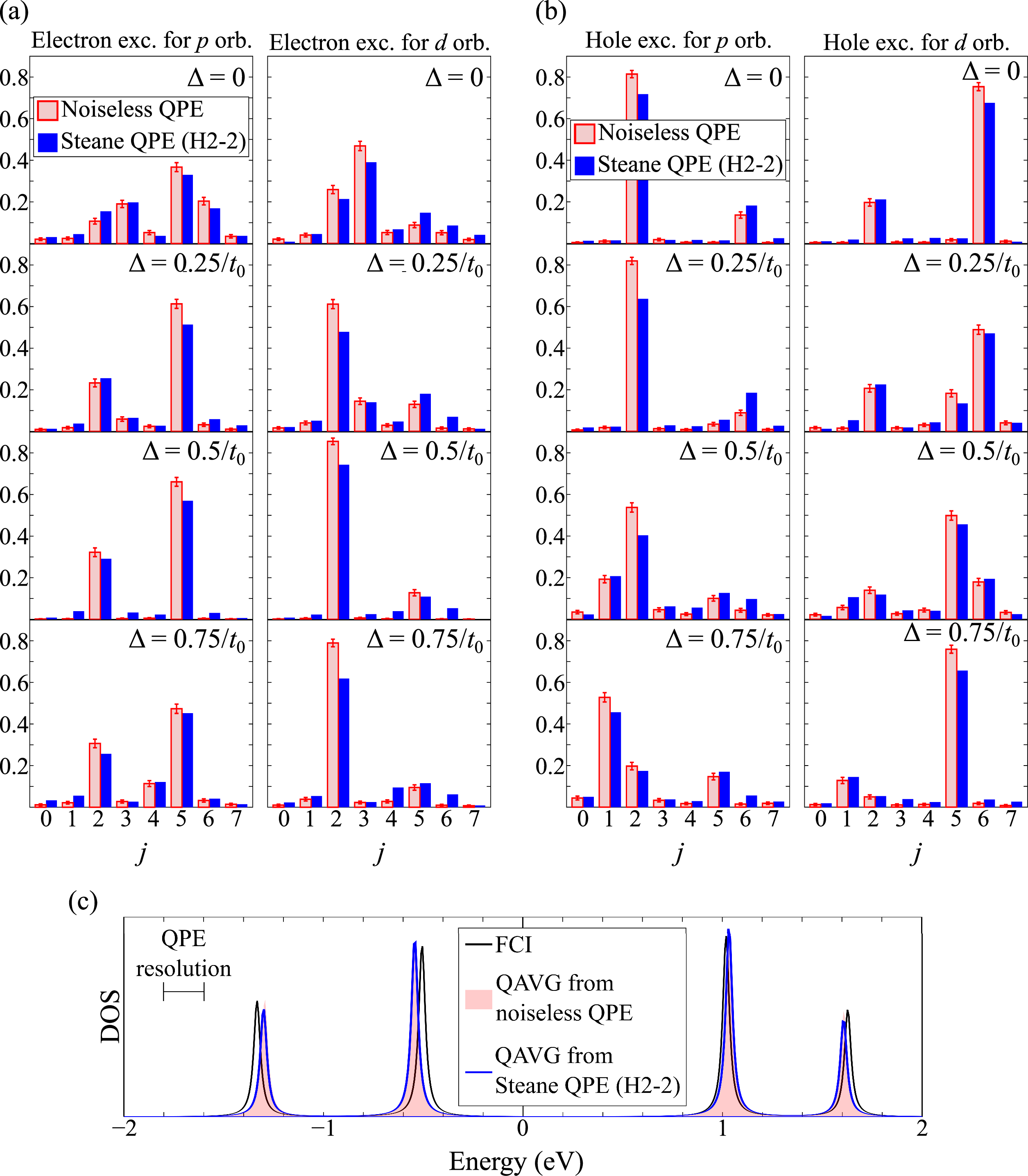}
\end{center}
\caption{
(a)
Normalized histograms obtained from the logical QPE circuit $\mathcal{C}^{\mathrm{(log, 1a)}}$ for electron excitations with four origin shifts $\Delta$ executed on H2-2.
For each of the histograms, $n_{\mathrm{shots}} = 500$ shots were performed.
$j$ represents each measurement outcome indicating the observed effective three ancillae.
The probability distributions $\mathbb{P}_j$ calculated from noiseless circuit simulations are also shown.
(b)
Histograms and related quantities obtained for hole excitations.
(c)
Many-electron DOS obtained from the QAVG calculations using the histograms with the four shifts.
Those from the FCI results and the noiseless circuit simulations are also shown.
}
\label{fig:qavg_logical_qpe}
\end{figure*}

We first collected histograms from logical QPE based on the Steane code on H2-2.
Specifically, we executed the circuits $\mathcal{C}^{\mathrm{(log, 1a)}}$
in Fig.~\ref{fig:qpe_circuits}(c), that implement the effective three-ancilla QPE.
We set the energy origin for the QPE circuits similarly to the physical circuits executed above.
For each of the histograms, we accumulated the measurement outcomes for $n_{\mathrm{shots}} = 500$ shots.
Figure \ref{fig:qavg_logical_qpe}(a) and (b) show the normalized histograms for the electron and hole excitations, respectively.
It is noted here that each of the histograms contains the results of shots fewer than $n_{\mathrm{shots}}$ since some of them were discarded due to the FT encoding.
Although the agreement between the histograms collected from the real quantum computer
and those from the noiseless simulations is reasonable,
it is not as good as the physical cases in Figs.~\ref{fig:qavg_fci_and_phys_qpe}(a) and (b).
It is because the noises on the logical QPE circuits,
that involved the more constituent physical qubits and the more physical gate operations than the physical QPE circuits,
deteriorated the results directly since the logical rotations were non-FT and we did not apply online error correction to the encoded states.

\subsubsection{QAVG calculations}

Having collected the histograms from $\mathcal{C}^{\mathrm{(log, 1a)}}$,
we minimized the cost values similarly to the cases of physical QPE.
The optimized trial parameters are provided in Table \ref{tab:qavg_optimized_params}.
We then reconstructed the GF from the optimized trial parameters to get the many-electron DOS,
as plotted in Fig.~\ref{fig:qavg_logical_qpe}(c).
It is surprising to see that the logical QPE provides the DOS in very good agreement with the noiseless DOS as well as the physical QPE in Fig.~\ref{fig:qavg_fci_and_phys_qpe}(c) despite the noisy histograms from $\mathcal{C}^{\mathrm{(log, 1a)}}.$

\begin{figure*}
\begin{center}
\includegraphics[width=17cm]{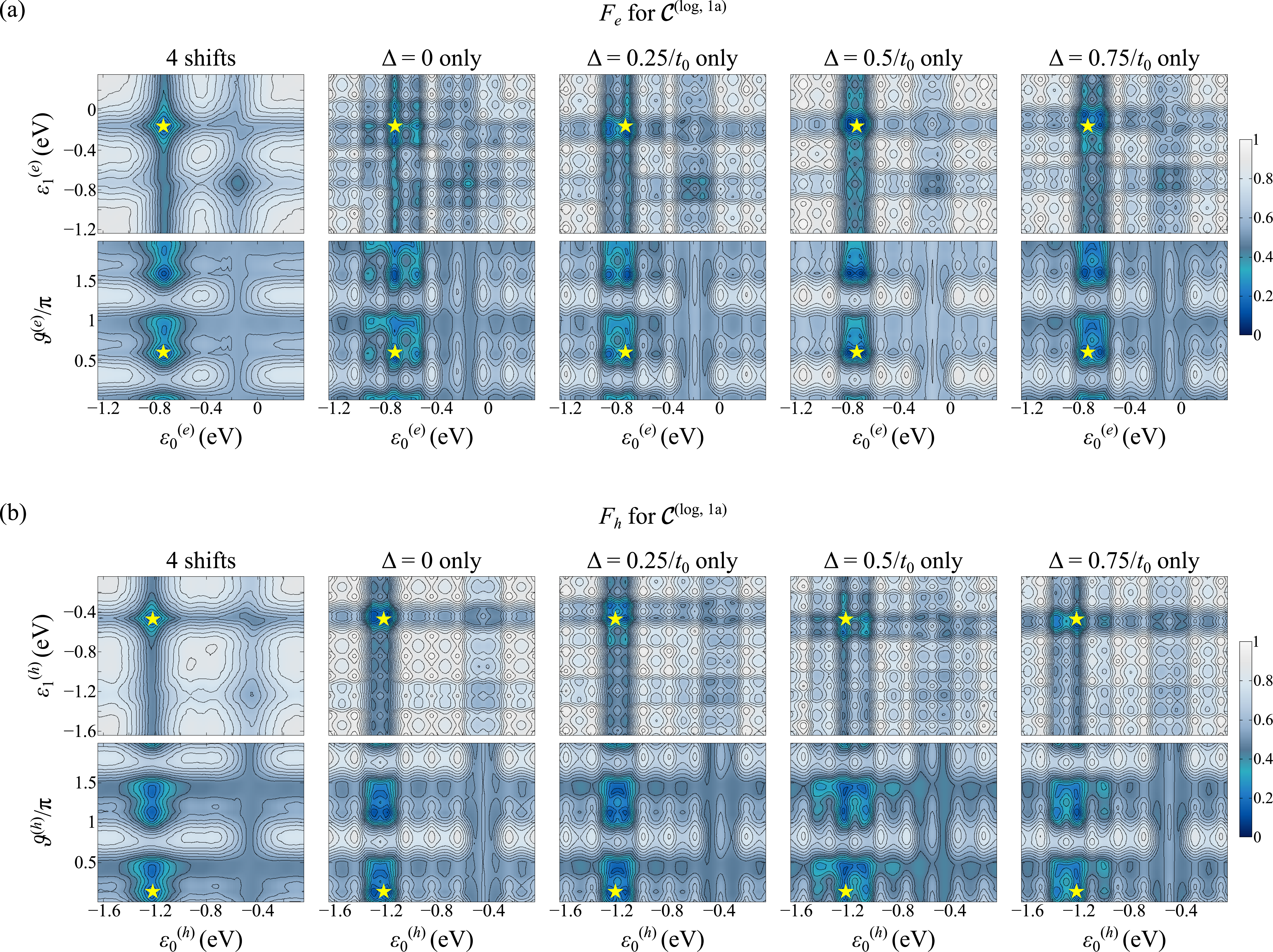}
\end{center}
\caption{
(a)
Cost landscapes for the electron excitations based on the histograms collected from the logical QPE circuits $\mathcal{C}^{\mathrm{(log, 1a)}}$ on H2-2.
The upper panels represent $F_e (\vartheta^{(e)}_{\mathrm{opt}}, \varepsilon^{(e)}_0, \varepsilon^{(e)}_1),$
while the lower panels represent
$F_e (\vartheta^{(e)}, \varepsilon^{(e)}_0, \varepsilon^{(e)}_{1, \mathrm{opt}}).$
The star symbols designate the optimized parameters $\Lambda^{(e)}_{\mathrm{opt}}$ for $\mathcal{C}^{\mathrm{(log, 1a)}}$ in Table \ref{tab:qavg_optimized_params}.
(b) Similar plots for the hole excitations.
}
\label{fig:qavg_cost_for_logical_qpe}
\end{figure*}

The cost landscapes for the logical QPE are plotted in Fig.~\ref{fig:qavg_cost_for_logical_qpe}.
We can see that the averaging over the origin shifts suppresses the local minima in the single-reference cost landscapes drastically as well as in the cases of physical QPE.
These results from the logical circuits suggest that physically motivated continuous parametrization and averaging over multiple QPE settings may render the determination of spectra robust against noises for a generic system.

\subsubsection{Survival history}

\begin{figure}
\begin{center}
\includegraphics[width=8cm]{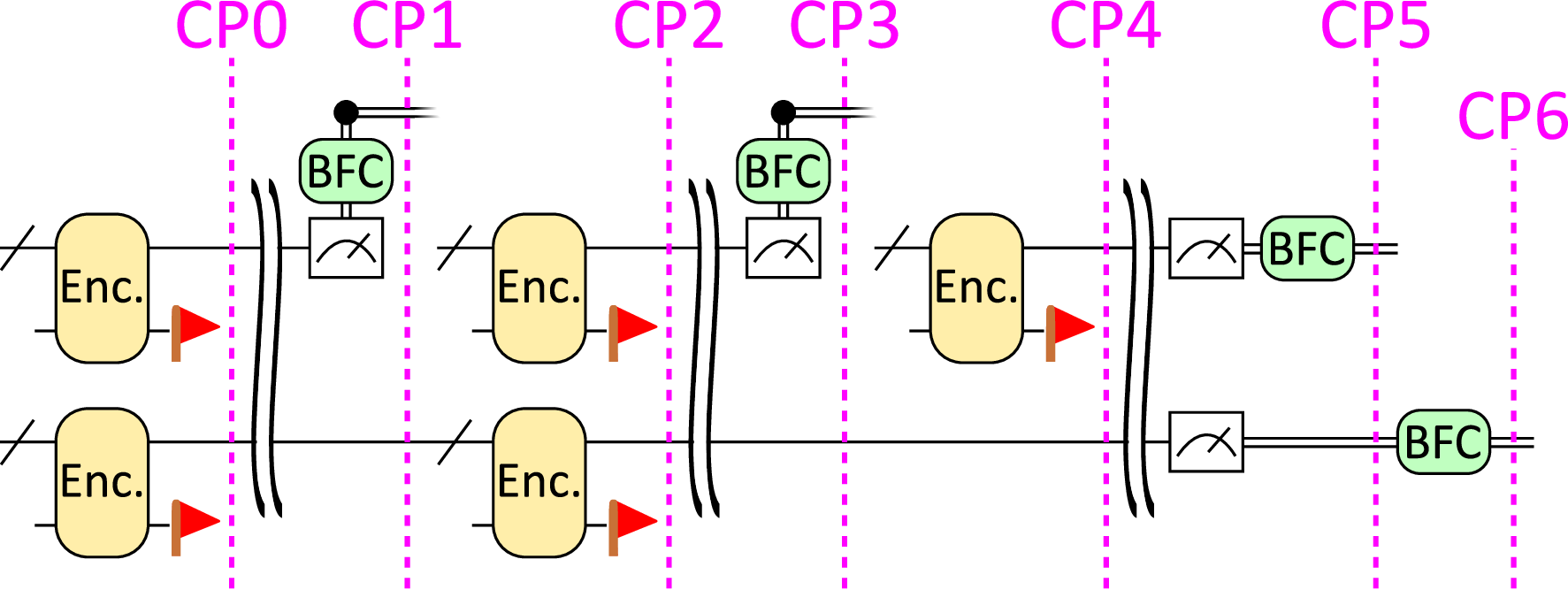}
\end{center}
\caption{
Definition of the checkpoints in $\mathcal{C}^{\mathrm{(log, 1a)}}.$
}
\label{fig:checkpoints_in_qpe_circuit}
\end{figure}

Since we executed 500 shots on the QPE circuit for each combination of the excitation ($e$ or $h$), the orbital ($p$ or $d$), and the shift (among four),
the total number of shots was 8000.
We found that the total discard ratio was $0.029,$ where 233 shots were discarded due to the harmful errors detected by the flag qubits.
To examine where the shots were discarded and corrected in detail,
we define the checkpoints in $\mathcal{C}^{(\mathrm{phys,1a})}$ as shown in 
Fig.~\ref{fig:checkpoints_in_qpe_circuit},
for which we provide the survival history of the shots in Table \ref{tab:survived_shots_in_steane}.
We see that the initialized and reset physical qubits were discarded just after encoding with a probability $\approx 1 \%$ at the checkpoints CP0, CP2, and the CP4.
As for BFC, the observed logical ancilla underwent the correction in each round of QFT with a probability $\approx 10 \%$ at CP1, CP3, and CP5.
The system logical qubit underwent the correction with a higher probability at CP6 since it had only a single chance to be corrected until the circuit was finished.

\begin{table*}[]
\centering
\caption{Survival history of shots performed on the logical QPE circuits $\mathcal{C}^{\mathrm{(log, 1a)}}$ on H2-2. This table provides the information about how many shots passed or were discarded or underwent BFC at the checkpoints defined in Fig.~\ref{fig:checkpoints_in_qpe_circuit}. The local discard at each checkpoint represents the number of shots discarded immediately before the checkpoint.}
\label{tab:survived_shots_in_steane}
\begin{tabular}{cllcc}
\hline
Checkpoint      & \multicolumn{1}{c}{\begin{tabular}[c]{@{}c@{}}Local discard\\ (local ratio)\end{tabular}} & \multicolumn{1}{c}{\begin{tabular}[c]{@{}c@{}}Accumulated discard\\ (accumulated ratio)\end{tabular}} & \begin{tabular}[c]{@{}c@{}}Correction\\ (local ratio)\end{tabular} & Survived \\ \hline
Iinitialization &                                                                                           &                                                                                                       & \multicolumn{1}{l}{}                                               & $8000$   \\
CP0             & \multicolumn{1}{c}{$114 \ (0.0143)$}                                                      & \multicolumn{1}{c}{$114 \ (0.0143)$}                                                                  & \multicolumn{1}{l}{}                                               & $7886$   \\
CP1             &                                                                                           &                                                                                                       & $731 \ (0.0927)$                                                   & $7886$   \\
CP2             & \multicolumn{1}{c}{$57 \ (0.0072)$}                                                       & \multicolumn{1}{c}{$169 \ (0.0211)$}                                                                  & \multicolumn{1}{l}{}                                               & $7829$   \\
CP3             &                                                                                           &                                                                                                       & $541 \ (0.0691)$                                                   & $7829$   \\
CP4             & \multicolumn{1}{c}{$64 \ (0.0082)$}                                                       & \multicolumn{1}{c}{$233 \ (0.0291)$}                                                                  & \multicolumn{1}{l}{}                                               & $7765$   \\
CP5             &                                                                                           &                                                                                                       & $1024 \ (0.1319)$                                                  & $7765$   \\
CP6             &                                                                                           &                                                                                                       & $1338 \ (0.1723)$                                                  & $7765$   \\ \hline
\end{tabular}
\end{table*}

\section{Conclusions}

We proposed the QAVG approach for finding the one-particle GF of a many-electron system based on QPE circuits specified by vernier parameters.
The combination of continuous parametrization and averaging enabled us to reconstruct the GF of the dimer model,
that reproduced the FCI-GF rather accurately.
The discrepancies between the QAVG and FCI results were found to be quite small compared to the QPE resolution not only for the physical circuits,
but also for the logical circuits based on the Steane code.
In addition, it was demonstrated that the averaging of cost suppresses the local minima of the cost landscapes.
This feature of the QAVG approach will be more prominent for larger systems since it provides a way for alleviating the curse of dimensionality.
Although the logical circuits in the present study achieved the satisfactory reproduction of the spectra,
the fidelity of QPE may be improved to some extent by exploiting syndrome extraction followed by discard or online error correction.
Since the approach was found to be robust against the noisy histograms from the logical circuits,
it will be useful in the early-FTQC era where the capability of quantum hardware has not achieved the requirements for genuine FTQC yet.

Although we adopted the dimer model for the real quantum computer in the present study,
we will be able to adopt larger and more realistic models after the advent of FTQC era in the future,
where hundreds of high-fidelity logical qubits are available.
In fact, we confirmed that 17 MLWOs can be straightforwardly constructed that originate from the two $2 \pi^*$ MOs and the 15 $d$ orbitals at the three Fe atoms bonded to the CO molecule.
Quantum computation for such a model or larger ones will provide spectra that can be compared directly with experimental results.

\begin{acknowledgments}
This work was partially supported by the Center of Innovations for Sustainable Quantum AI (JST Grant Number JPMJPF2221).
The computation in this work has been done using the facilities of the Supercomputer Center, the Institute for Solid State Physics, the University of Tokyo (ISSPkyodo-SC-2026-Ea-0014).
\end{acknowledgments}

\begin{widetext}

\appendix

\section{Numerical FCI results for dimer model}
\label{sec:FCI_results}

\subsection{Ground state}

We performed FCI calculations for the dimer model with the parameters shown in Table \ref{tab:hamiltonian_params} for the chemical-potential shift $\Delta \mu = 1.5$ eV.
The ground state $| \Psi_{\mathrm{gs}} \rangle$ over all the possible combinations of the number $n_e$ of electrons and the $z$ component $S_z$ of the total spin was found to live in the sector $(n_e = 2, S_z = 0).$
Its energy eigenvalue is $E_{\mathrm{gs}} = -1.76303$ eV.
The normalized ground state is expressed as a vector in this four-dimensional subspace as
\begin{align}
    | \Psi_{\mathrm{gs}} \rangle
    =
        0.09776
        a_{p \uparrow}^\dagger a_{p \downarrow}^\dagger | \mathrm{vac} \rangle
        +
        0.69307
        a_{d \uparrow}^\dagger a_{p \downarrow}^\dagger | \mathrm{vac} \rangle
        +
        0.69307
        a_{p \uparrow}^\dagger a_{d \downarrow}^\dagger | \mathrm{vac} \rangle
        +
        0.17249
        a_{d \uparrow}^\dagger a_{d \downarrow}^\dagger | \mathrm{vac} \rangle
        .
        \label{dimer_FCI:ground_state}
\end{align}
The one-electron density matrix for the spin-up sectors is thus
\begin{align}
    \gamma_{\uparrow}
    =
        \begin{pmatrix}
            \gamma_{\uparrow p p} & \gamma_{\uparrow p d} \\
            \gamma_{\uparrow d p} & \gamma_{\uparrow d d}
        \end{pmatrix}
    =
        \begin{pmatrix}
            \langle \Psi_{\mathrm{gs}} | a_{p \uparrow}^\dagger a_{p \uparrow} | \Psi_{\mathrm{gs}} \rangle
            &
            \langle \Psi_{\mathrm{gs}} | a_{p \uparrow}^\dagger a_{d \uparrow} | \Psi_{\mathrm{gs}} \rangle
            \\
            \langle \Psi_{\mathrm{gs}} | a_{d \uparrow}^\dagger a_{p \uparrow} | \Psi_{\mathrm{gs}} \rangle
            &
            \langle \Psi_{\mathrm{gs}} | a_{d \uparrow}^\dagger a_{d \uparrow} | \Psi_{\mathrm{gs}} \rangle
        \end{pmatrix}
    =
        \begin{pmatrix}
            0.48990 & 0.18730 \\
            0.18730 & 0.51010
        \end{pmatrix}
        ,
\end{align}
which is equal to that for the spin-down sectors since the ground state is spin unpolarized:
$\gamma_\downarrow = \gamma_\uparrow.$

\subsection{Electron-excited states}

From Eq.~(\ref{dimer_FCI:ground_state}),
the unnormalized electron-excited state for a spin-up $p$ electron is given by
\begin{align}
    a_{p \uparrow}^\dagger
    | \Psi_{\mathrm{gs}} \rangle
    =
        0.69307
        a_{p \uparrow}^\dagger
        a_{d \uparrow}^\dagger a_{p \downarrow}^\dagger | \mathrm{vac} \rangle
        +
        0.17249
        a_{p \uparrow}^\dagger
        a_{d \uparrow}^\dagger a_{d \downarrow}^\dagger | \mathrm{vac} \rangle
    ,
\end{align}
while that for a spin-up $d$ electron is given by
\begin{align}
    a_{d \uparrow}^\dagger
    | \Psi_{\mathrm{gs}} \rangle
    =
        -
        0.09776
        a_{p \uparrow}^\dagger
        a_{d \uparrow}^\dagger
        a_{p \downarrow}^\dagger | \mathrm{vac} \rangle
        -
        0.69307
        a_{p \uparrow}^\dagger
        a_{d \uparrow}^\dagger
        a_{d \downarrow}^\dagger | \mathrm{vac} \rangle
    .
\end{align}
These excited states live in the two-dimensional sector $(n_e = 3, S_z = 1/2).$
As explained in the main text, we map the two bases
$
a_{p \uparrow}^\dagger
a_{d \uparrow}^\dagger
a_{p \downarrow}^\dagger | \mathrm{vac} \rangle
$
and
$
a_{p \uparrow}^\dagger
a_{d \uparrow}^\dagger
a_{d \downarrow}^\dagger | \mathrm{vac} \rangle
$
to single-qubit states $| 0 \rangle$ and $| 1 \rangle,$ respectively, for the QPE circuit.
$a_{p \uparrow}^\dagger | \Psi_{\mathrm{gs}} \rangle$ (multiplied by a normalization factor) can thus be prepared by applying to an initialized qubit a $y$ rotation gate $R_y (2 \eta_{\mathrm{exc}})$ with the angle $\eta_{\mathrm{exc}} \equiv 0.24392,$
while $a_{d \uparrow}^\dagger | \Psi_{\mathrm{gs}} \rangle$ can be prepared by 
$R_y (2 \zeta_{\mathrm{exc}})$ with $\zeta_{\mathrm{exc}} \equiv 1.43067.$

\subsection{Hole-excited states}

From Eq.~(\ref{dimer_FCI:ground_state}),
the unnormalized hole-excited state for a spin-up $p$ electron is given by
\begin{align}
    a_{p \uparrow}
    | \Psi_{\mathrm{gs}} \rangle
    =
        0.09776
        a_{p \downarrow}^\dagger | \mathrm{vac} \rangle
        +
        0.69307
        a_{d \downarrow}^\dagger | \mathrm{vac} \rangle
        ,
\end{align}
while that for a spin-up $d$ electron is given by
\begin{align}
    a_{d \uparrow}
    | \Psi_{\mathrm{gs}} \rangle
    =
        0.69307
        a_{p \downarrow}^\dagger | \mathrm{vac} \rangle
        +
        0.17249
        a_{d \downarrow}^\dagger | \mathrm{vac} \rangle
        .
\end{align}
These excited states live in the two-dimensional sector $(n_e = 1, S_z = -1/2).$
As explained in the main text, we map the two bases
$a_{p \downarrow}^\dagger | \mathrm{vac} \rangle$ and
$a_{d \downarrow}^\dagger | \mathrm{vac} \rangle$
to single-qubit states $| 0 \rangle$ and $| 1 \rangle,$ respectively, for the QPE circuit.
$a_{p \uparrow} | \Psi_{\mathrm{gs}} \rangle$
and
$a_{d \uparrow} | \Psi_{\mathrm{gs}} \rangle$
can thus be prepared by the rotation gates $R_y (2 \zeta_{\mathrm{exc}})$ and $R_y (2 \eta_{\mathrm{exc}}),$ respectively.

\section{Expressions for $\mathbb{P}_\lambda^{(\kappa \uparrow, \xi)} (\vartheta^{(\xi)})$}
\label{sec:expr_for_excitation_probs}

By defining
\begin{align}
    S_{\kappa \uparrow}^{(\nu \nu', \xi)}
    \equiv
        \begin{cases}
                d_{\kappa \uparrow}^{(\nu, e) *}
                d_{\kappa \uparrow}^{(\nu', e) }
            /(1 - \gamma_{\uparrow \kappa \kappa} ) & \xi = e
            \\
                d_{\kappa \uparrow}^{(\nu, h) *}
                d_{\kappa \uparrow}^{(\nu', h)}
            /\gamma_{\uparrow \kappa \kappa} & \xi = h
        \end{cases}
    ,
\end{align}
we can rewrite
Eqs.~(\ref{prob_for_obs_exc_state_e}) and (\ref{prob_for_obs_exc_state_h}) as
$
\mathbb{P}_\lambda^{(\kappa \uparrow, \xi)}
=
\sum_{\nu, \nu'}
S_{\kappa \uparrow}^{(\nu \nu', \xi)}
\widetilde{b}_{\nu \uparrow}^{(\lambda, \xi) *}
\widetilde{b}_{\nu' \uparrow}^{(\lambda, \xi)}
.
$
We replace $\{ \widetilde{b}_{\nu \uparrow}^{(\lambda, \xi)} \}_{\nu, \lambda, \xi}$
with the real trial parameters $\{ v_{\nu}^{(\lambda, \xi)} \}_{\nu, \lambda, \xi},$
which we parametrize with $\{ \vartheta^{(\xi)} \}_\xi,$ as illustrated in the main text.
For the parametrization in Eq.~(\ref{param_v_using_vartheta}),
the theoretical probability distributions for the excited states $\lambda = 0$ and $1$
appearing in Eq.~(\ref{prob_distr_for_traial_params}) are thus given by
\begin{align}
    \mathbb{P}_{\lambda = 0}^{(\kappa \uparrow, \xi)} (\vartheta^{(\xi)})
    =
        S_{\kappa \uparrow}^{(00, \xi)}
        \cos^2 \vartheta^{(\xi)}
        -
        \left(
            S_{\kappa \uparrow}^{(01, \xi)}
            +
            S_{\kappa \uparrow}^{(10, \xi)}
        \right)
        \cos \vartheta^{(\xi)}
        \sin \vartheta^{(\xi)}
        +
        S_{\kappa \uparrow}^{(11, \xi)}
        \sin^2 \vartheta^{(\xi)}
\end{align}
and
\begin{align}
    \mathbb{P}_{\lambda = 1}^{(\kappa \uparrow, \xi)} (\vartheta^{(\xi)})
    =
        S_{\kappa \uparrow}^{(00, \xi)}
        \sin^2 \vartheta^{(\xi)}
        +
        \left(
            S_{\kappa \uparrow}^{(01, \xi)}
            +
            S_{\kappa \uparrow}^{(10, \xi)}
        \right)
        \cos \vartheta^{(\xi)}
        \sin \vartheta^{(\xi)}
        +
        S_{\kappa \uparrow}^{(11, \xi)}
        \cos^2 \vartheta^{(\xi)}
    ,
\end{align}
respectively.

\end{widetext}

\bibliography{ref}

\end{document}